\documentclass[aps,prd,twocolumn,nofootinbib,groupedaddress,amsfonts,floatfix]{revtex4}

\usepackage{setspace}
\usepackage{amsmath}
\usepackage{graphicx}
\usepackage{color}
\usepackage{array}

\makeatletter
\newcommand{\thickhline}{%
    \noalign {\ifnum 0=`}\fi \hrule height 1.5pt
    \futurelet \reserved@a \@xhline
}
\newcolumntype{"}{@{\vrule width 1.5pt}}
\makeatother

\begin{document}

\title{Advanced LIGO's ability to detect apparent violations of the cosmic censorship conjecture and the no-hair theorem through compact binary coalescence detections}
\author{Madeline Wade${}^1$}
\author{Jolien D.E. Creighton${}^1$}
\author{Evan Ochsner${}^1$}
\author{Alex B. Nielsen${}^2$}
\affiliation{${}^1$Department of Physics, University of Wisconsin -- Milwaukee, CGCA, P.O. Box 413, Milwaukee, Wisconsin 53201 \\
${}^2$Max-Planck-Institute for Gravitational Physics, 38 Callinstrasse, 30167 Hanover, Germany}

\begin{abstract}
We study the ability of the Advanced Laser Interferometer Gravitational-Wave Observatory (aLIGO) to detect apparent violations of the cosmic censorship conjecture and the no-hair theorem.  The cosmic censorship conjecture, which is believed to be true in the theory of general relativity, limits the spin-to-mass-squared ratio of a Kerr black hole, $\chi \equiv j/m^2 \le 1$.  The no-hair theorem, which is also believed to be true in the theory of general relativity, suggests a particular value for the tidal Love number of a nonrotating black hole ($k_2=0$).  Using the Fisher matrix formalism, we examine the measurability of the spin and tidal deformability of compact binary systems involving at least one putative black hole.  Using parameter measurement errors and correlations obtained from the Fisher matrix, we determine the smallest detectable violation of bounds implied by the cosmic censorship conjecture and the no-hair theorem.  We examine the effect of excluding unphysical areas of parameter space when determining the smallest detectable apparent violations, and we examine the effect of different post-Newtonian corrections to the amplitude of the compact binary coalescence gravitational waveform, as given in Arun {\it et al.} [Phys. Rev. D 79, 104023 (2009)].  In addition, we perform a brief study of how the recently calculated 3.0 pN and 3.5 pN spin-orbit corrections to the phase [Marset {\it et al.} Classical Quantum Gravity 30, 055007 (2013)] affect spin and mass parameter measurability.  We find that physical priors on the symmetric mass ratio and higher harmonics in the gravitational waveform could significantly affect the ability of aLIGO to investigate cosmic censorship and the no-hair theorem for certain systems.
\end{abstract}

\maketitle

\section{Motivation}
\label{sec:intro}

The era of advanced gravitational-wave detectors is expected to provide the first direct observations of gravitational waves.  The inspiral portion of compact binary coalescence (CBC) events are the most promising sources for gravitational-wave detections in ground-based interferometers, such as the Advanced Laser Interferometer Gravitational-Wave Observatory (aLIGO).  Expected detection rates for binary black hole (BBH) mergers range from 0.4 to 1000 per year with a realistic rate of 20 per year, and expected detection rates for neutron-star--black-hole (NS-BH) mergers range from 0.2 to 300 per year with a realistic rate of 10 per year \cite{rates}.  The form of the gravitational-wave strain depends on the chosen metric theory of gravity.  The most accepted theory of gravity is Einstein's theory of general relativity.  An important use of gravitational-wave detectors will be to test the theory of general relativity and cosmological conjectures associated with general relativity.  

Even within the confines of general relativity, there are conjectures that, while widely believed, have not been absolutely established, and violations could be uncovered by gravitational-wave observations.  One such conjecture that is believed to be true in general relativity is the cosmic censorship conjecture, which states roughly that all singularities in spacetime must have an event horizon that conceals the singularity from a distant observer \cite{cosmiccensorship}.  In the Kerr geometry of a spinning black hole, the event horizon can only exist for mass and spin ratios that satisfy the Kerr bound, $j \le m^2$ in geometric units (adopted throughout this paper), where $j$ is the spin of the black hole and $m$ is the mass of the black hole.  If the spin of a compact object exceeds the value of its mass squared, then the compact object violates the cosmic censorship conjecture within the context of the Kerr geometry \cite{Kerr, MTW, Hartle}.  This limit is often expressed in terms of the Kerr parameter $\chi \equiv j/m^2 \le 1$.

The no-hair theorem is a consequence of the theory of general relativity. The no-hair theorem states that a regular black hole that has settled down to its final stationary vacuum state is determined only by its mass, spin and electric charge \cite{nohair1, nohair2, nohair3, nohair4, nohair5, nohair6, MTW, lrr-2012-7}.  Astrophysical black holes are thought to be electrically neutral, and therefore would be categorized just by their mass and spin.  It is widely expected that black holes in binary systems will be closely described by such simple states for most of the inspiral phase.  Although the black hole will be slightly tidally distorted by its binary partner, it has been shown that the relativistic tidal Love number of a nonrotating black hole will still be zero \cite{Binnington}.  While nothing in the literature shows that the tidal Love number should be zero for rotating black holes, we suspect it should still be small for this scenario.  Thus if the post-Newtonian tidal Love number is found to deviate from zero for a nonrotating object, it can be seen as evidence that the requirements of the no-hair theorem are not fulfilled, since the black hole is no longer uniquely defined by its mass, spin and electric charge.  If the object is too massive to be a neutron star (i.e. $m_i > 3$ ${\rm M}_\odot$)\footnote{Reasonably general arguments show that compact objects having $m > 3$ ${\rm M}_\odot$ should be fully-collapsed black holes \cite{Duncan}.  Though it is possible that exotic objects will have masses with $m>3$ ${\rm M}_\odot$.}, then it is likely to be some exotic object far from the Schwarzschild solution. A more detailed discussion of the implications of the no-hair theorem can be found in Sec.~\ref{sec:tidalparams}.

The gravitational-wave strain produced by the inspiral portion of a CBC event depends on the system's parameters, such as component masses, component spins, and component tidal Love numbers.  Once a gravitational-wave detection is made by aLIGO, parameter estimation techniques will be used to extract the system's most likely parameters from the raw data.  This will be done using full Bayesian analyses that involve techniques such as Markov-chain Monte Carlo and nested sampling.  An in-depth discussion of LIGO parameter estimation can be found in Ref. \cite{PEPaper}.  Based on the results of parameter estimation, if at least one of the system's measured component masses indicates that a body should nominally be a black hole, then the system can be used to test for apparent violations of the cosmic censorship conjecture and the no-hair theorem.

Many other authors have investigated the possibility of using gravitational-wave observations to test aspects of general relativity.  These include measuring the deviation of post-Newtonian coefficients from their predicted values in general relativity \cite{Arun:2006hn,Yunes:2009ke,Li:2011cg}, looking for alternative wave-polarization states that do not occur in general relativity \cite{Chatziioannou:2012rf,Hayama:2012au}, testing for a nonzero graviton mass \cite{Arun:2009pq,Mirshekari:2011yq,Keppel:2010qu}, and exploring whether the ringdown signal is consistent with the quasinormal modes of a Kerr black hole \cite{Gossan:2011ha,Dreyer:2003bv,Kamaretsos:2011um}.  For recent reviews of these techniques, see Ref. \cite{Broeck:2013kx,Arun:2013bp,Yunes:2013dva}.  Rodriguez, Mandel and Gair look at aLIGO's ability to verify the no-hair theorem for intermediate-mass black hole systems in Ref. \cite{Carl}.  Tests of the no-hair theorem and cosmic censorship can also be conducted in the electromagnetic sector using a variety of techniques, including accretion disk modeling \cite{Johannsen:2012ng}, observations of orbiting stars and gas \cite{Sadeghian:2011ub}, and pulsar orbit timing \cite{Stairs:2003eg}.

This paper is outlined in the following manner: In Sec.~\ref{sec:background} we provide background information pertinent to our studies.  In Sec.~\ref{sec:waveforms} we outline the gravitational waveform for CBC events. In Sec.~\ref{sec:fishersvd} we describe the Fisher matrix formalism, discuss the validity of the Fisher matrix approach, and describe a singular-value decomposition method that we use to assist in inverting the Fisher matrix.  In Sec.~\ref{sec:params} we discuss the parameters used in each gravitational waveform, and we outline known bounds on the chosen parameter space.  In Sec.~\ref{sec:results}, we discuss our results for the ability of aLIGO to detect apparent violations of the cosmic censorship conjecture and the no-hair theorem.  In Sec.~\ref{sec:discussion} we discuss future implications of our findings.  In Sec.~\ref{sec:conclusions} we succinctly summarize our results.

(Units convention: $G=c=1$.)

\section{Background}
\label{sec:background}

The inspiral portion of the CBC gravitational waveform is well modeled by post-Newtonian (pN) expansions to the phase and amplitude of the waveform \cite{1938, Epstein, MTW, Weinberg, Blanchet, Jolien, BIOPS}.  However, systematic biases due to the deviation of a post-Newtonian waveform from the true waveform can significantly affect parameter estimation.  Therefore, when using post-Newtonian waveforms, it is important to employ the most up-to-date and accurate calculations.  In this paper, we use the waveforms provided in Arun {\it et al.} \cite{Evan} that include post-Newtonian expansions of the phase  to 3.5 pN order and of the amplitude to 2.5 pN order.  Spin corrections are calculated for both the post-Newtonian phase to 2.5 pN order and amplitude to 2.0 pN order in Ref.~\cite{Evan}.  When this work was near completion, the 3.0 pN- and 3.5 pN-order spin-orbit phase corrections were calculated by Marsat {\it et al.} in Ref.~\cite{Bohe}.  We briefly investigate how these affect our results.  For nonspinning, tidal waveforms, we use the lowest-order tidal correction to the phase of the waveform given in Ref.~\cite{Ben}.

To estimate the measurability of parameters appearing in the inspiral CBC gravitational waveform, we use the Fisher matrix formalism for a single detector, described in Sec.~\ref{sec:fishersvd}.  The accuracy of measurement errors produced by the Fisher matrix formalism is a function of the signal-to-noise ratio (SNR).  A noisier system will bias parameter error estimates obtained with the Fisher matrix formalism \cite{Michele}.  However, a more accurate Bayesian approach to parameter estimation using techniques such as Markov-chain Monte Carlo (MCMC) can be very computationally expensive.  For the purpose of preliminary investigation, the Fisher matrix formalism does well to indicate the effects that should be studied more closely with a full Bayesian analysis.  

Much work has been done on parameter measurability for CBC systems using the Fisher matrix formalism and post-Newtonian CBC inspiral waveforms.  Cutler and Flanagan \cite{Cutler} studied the measurability of spin parameters, along with other parameters, for the gravitational waveform with a Newtonian-amplitude (0.0 pN-order correction to the amplitude) and 1.5 pN-order corrections to the phase.  Poisson and Will \cite{poissonandwill} and Krolak {\it et al.} \cite{Krolak} expanded the study for Newtonian-amplitude waveforms with 2.0 pN-order phase corrections.  Arun {\it et al.}  \cite{Arun} studied parameter estimation for nonspinning waveforms with a Newtonian-amplitude and phase corrections to 3.5 pN order.  Van Den Broeck and Sengupta \cite{Chris} included post-Newtonian corrections to the amplitude of the waveform and kept corrections to 3.5 pN order in the phase of the waveform, including spin effects in the phase.  Nielsen \cite{Alex} studied a Newtonian-amplitude waveform with additional spin-spin and spin-orbit corrections appearing in the phase of the waveform as derived in Ref.~\cite{Evan}.

In this work, we investigate aLIGO's ability to detect apparent violations of the cosmic censorship conjecture and the no-hair theorem.  We study how different post-Newtonian approximations to the amplitude of the gravitational waveform affect parameter measurability.  We include the post-Newtonian phase corrections to 3.5 pN order with spin-orbit and spin-spin\footnote{The ``spin-spin'' corrections include not only $\vec\chi_1 \cdot \vec\chi_2$ corrections, but also quadrupole-monopole and the so-called ``self-spin'' terms $\propto \chi_1^2$.} phasing corrections to 2.5 pN and 2.0 pN order respectively, and we vary the post-Newtonian-amplitude corrections from 0.0 pN to 2.5 pN order.  We also study the effect of spin corrections in the amplitude of the waveform \cite{Evan} and of the recent spin-orbit corrections to the phase of the waveform at 3.0 pN and 3.5 pN order \cite{Bohe}.

We investigate how prior knowledge about unphysical areas of parameter space can affect the measurability of spin and tidal parameters appearing in the waveform.  We have not done this by incorporating a prior into the Fisher matrix calculations.   It is difficult to incorporate flat priors into the Fisher matrix formalism, although this has been studied by Nielsen \cite{Alex}.  Instead, in this work we restrict some of the parameter space after a full Fisher matrix calculation has been carried out.  

\section{Compact Binary Coalescence Gravitational Waveform}
\label{sec:waveforms}

The gravitational-wave strain for the inspiral portion of a compact binary coalescence event has the following general form in the frequency domain
\begin{equation}
\label{eq:simplewaveform}
\tilde h(f) = A(f; \vec{\theta}) e^{i \Psi(f; \vec{\theta})} \ ,
\end{equation}
where $f$ is the gravitational-wave frequency and $\vec{\theta}$ are the parameters of the system producing the gravitational-wave signal \cite{MTW}.  The amplitude $A$ and the phase $\Psi$ can be expanded in a post-Newtonian (pN) approximation, and the phase is found using the stationary phase approximation (SPA).  The form for the pN expanded waveform given in Ref.~\cite{Evan} is
\begin{equation}
\label{eq:waveform}
\tilde h(f) = \frac{M^2}{D_M} \sqrt{\frac{5 \pi \eta}{48}} \sum_{n=0}^{N} \sum_{k=0}^{K} v_k^{n-\frac{7}{2}} C_k^{(n)} e^{i[k\Psi_{\rm SPA}(v_k) - \pi/4]} \ ,
\end{equation}
where $M=m_1+m_2$ is the total post-Newtonian mass of the binary system, $D_M$ is the transverse comoving distance (see Ref.~\cite{Hogg}), $\eta = m_1m_2/M^2$ is the symmetric mass ratio, $\Psi_{\rm SPA}$ is the SPA for the phase of the waveform to some chosen pN order (see below), the index $n$ indicates twice the pN expansion order of the amplitude, $N$ is twice the highest included pN expansion order of the amplitude, the index $k$ indicates the $kth$ harmonic, $K$ is the highest included harmonic, and the $C_k^{(n)}$ coefficients are given in Appendix D of Ref.~\cite{Evan}.  The dimensionless pN expansion parameter $v_k$ for the $kth$ harmonic is
\begin{equation}
v_k = \left(2 \pi M \frac{f}{k}\right)^{1/3} \ .
\end{equation}
The gravitational-wave frequency $f$ is related to the orbital frequency $F$ of the binary system through $f = k F$.  

We restrict our studies to spin-aligned (or antialigned), nonprecessing systems, where the spin is defined in the standard post-Newtonian fashion.  In reality, precession should be included in the gravitational waveform model \cite{Kidder, ATCST}.  This is especially important for unequal-mass systems, such as NS-BH binaries.  The size of the precession cone scales with the mass ratio in such a way that unequal-mass systems will precess more than equal-mass systems.  The effect of precession on parameter estimation has been studied in depth for space-based detectors \cite{LangHughes, Vecchio, LHC}.  In these studies, it is found that precession improves parameter estimation by breaking parameter degeneracies, but astrophysical systems may not have enough precession to induce this effect.  There are fewer studies of parameter estimation that include precession for ground-based detectors.  The LIGO-Virgo Collaboration performed parameter estimation for a few precessing models in Ref.~\cite{PEPaper}.  The effect of precession upon detection, rather than parameter estimation, for ground-based interferometers was recently studied in Ref.~\cite{HNBLOK}.  Recent studies of precession for LIGO parameter estimation include Ref.~\cite{BLO, Cho2013, Pekowsky}, but there are no definitive conclusions on how precession will affect parameter estimation for ground-based detectors.  Large-scale, systematic Bayesian inference analyses will likely be required to develop a better understanding of how precession will impact parameter estimation in the aLIGO era.  For simplicity, we have not investigated precessing systems in this work.

We study waveforms with amplitude corrections up to the 2.5 pN order ($N=5$), which include up to seven harmonics ($K=7$) in the waveform.  Post-Newtonian corrections for spinning systems have been investigated at length in, for example, Refs.~\cite{Evan, Bohe, spinref1, spinref2, spinref3, spinref4, spinref5, spinref6, spinref7}.  We include spin corrections to amplitude and phase as found in Ref.~\cite{Evan}. These include spin-orbit corrections calculated at 1.5 pN and 2.5 pN order in the phase, spin-spin corrections at 2.0 pN order in the phase, spin-orbit corrections appearing at 1.0 pN and 1.5 pN order in the amplitude, and spin-spin corrections appearing at 2.0 pN order in the amplitude.  Separately, we also study spin-orbit corrections that appear at 3.0 pN and 3.5 pN order in the phase as recently calculated in Ref.~\cite{Bohe}.  We investigate both spinning waveforms with no tidal corrections and nonspinning waveforms with the leading-order tidal correction to the phase, which appears at 5.0 pN order.

\pagebreak

The general SPA phase $\Psi_{\rm SPA}$ used in Eq.~\eqref{eq:waveform} is
\begin{widetext}
\begin{eqnarray}
\nonumber
\Psi_{\rm SPA}(v_k) &=& \frac{v_k^3}{M}t_c - \phi_c + \frac{3}{256} \frac{1}{v_k^5 \eta} \left\{1+\left(\frac{3715}{756} + \frac{55}{9}\eta\right) v_k^2 + \left(4 \epsilon \beta - 16 \pi\right) v_k^3 \right . \\
\nonumber
&&+ \left(\frac{15293365}{508032} + \frac{27145}{504}\eta + \frac{3085}{72}\eta^2 - 10 \epsilon \sigma\right) v_k^4
 + \left(\frac{38645\pi}{756} - \frac{65\pi}{9}\eta - \epsilon \gamma\right)\left(1+3 \ln\left[\frac{v_k}{v_{\rm ref}}\right]\right) v_k^5 \\
\nonumber
&&+ \left[\frac{11583231236531}{4694215680} - \frac{6848}{21}  \gamma_{\rm E} - \frac{640 \pi^2}{3} + \left(\frac{2255 \pi^2}{12} - \frac{15737765635}{3048192}\right) \eta\right . \\
\nonumber
&&+ \frac{76055}{1728} \eta^2 - \frac{127825}{1296} \eta^3\left . - \frac{6848}{21} \ln\left(4 v_k\right) + \alpha\left(160 \pi \beta - 20 \xi\right) \right] v_k^6 \\
\nonumber
&& + \left[ \frac{77096675\pi}{254016} + \frac{378515\pi\eta }{1512} - \frac{74045\pi\eta^2}{756} + \alpha \left(-20 \zeta + \gamma \left(-\frac{2229}{112} - \frac{99\eta}{4}\right)  \right . \right .\\
\label{eq:spinphase}
&&\left .\left . \left .+  \beta \left(\frac{43939885}{254016}+\frac{259205\eta}{504} + \frac{10165 \eta^2}{36} \right)\right) \right] v_k^7 \right \} \ ,
\end{eqnarray}
\end{widetext}
where $\epsilon$ and $\alpha$ are either 1 or 0 to turn on or off spin corrections to the phase ($\epsilon$ turns on or off the 1.5 pN- to 2.5 pN-order corrections and $\alpha$ turns on or off the 3.0 pN- and 3.5 pN-order corrections), $t_c$ and $\phi_c$ are the time and phase of coalescence, $\mathcal{M} = M \eta^{3/5}$ is the chirp mass, $\gamma_{\rm E}= 0.577216...$ is Euler's constant, and $v_{\rm ref}$ is an integration constant, which we take to equal 1.

The five spin parameters  appearing in $\Psi_{\rm SPA}$ and derived in Refs.~\cite{Evan, Bohe}--$\beta$, $\sigma$, $\gamma$, $\xi$, and $\zeta$--are
\begin{eqnarray*}
\beta &=& \sum_{i=1}^2\left(\frac{113}{12} \left(\frac{m_i}{M}\right)^2 + \frac{25}{4}\eta\right) \vec \chi_i \cdot \hat{{\rm L}}_{\rm N} \ , \\
\nonumber
\sigma &=& \eta \left[\frac{721}{48} \left(\vec \chi_1 \cdot \hat{{\rm L}}_{\rm N}\right) \left(\vec \chi_2 \cdot \hat{\rm L}_{\rm N} \right) - \frac{247}{48}\left( \vec \chi_1 \cdot \vec \chi_2\right) \right ] \\
\nonumber
&& \sum_{i=1}^2 \left \{\frac{5}{2} q_i \left(\frac{m_i}{M}\right)^2 \left[3 \left(\vec \chi_i \cdot \hat{\rm L}_{\rm N}\right)^2 - \chi_i^2\right] \right . \\
&&\left .+ \frac{1}{96} \left(\frac{m_i}{M}\right)^2\left[7  \chi_i^2 - \left( \vec \chi_i \cdot \hat{\rm L}_{\rm N}\right)^2\right]\right \} \ , \\
\nonumber
\gamma &=& \sum_{i=1}^2 \left [ \left ( \frac{732985}{2268} + \frac{140}{9} \eta\right)\left(\frac{m_i}{M}\right)^2 \right . \\
&&\left. + \eta \left(\frac{13915}{84} - \frac{10}{3} \eta\right)\right]\vec \chi_i \cdot \hat{\rm L}_{\rm N} \ , \\
\xi &=& \sum_{\i=1}^2 \left[ \frac{75 \pi}{2} \left(\frac{m_i}{M}\right)^2 + \frac{151 \pi}{6} \eta\right] \vec \chi_i \cdot \hat{\rm L}_{\rm N}  \ , \\
\nonumber
\zeta &=& \sum_{i=1}^2 \left[ \left(\frac{m_i}{M}\right)^2 \left (\frac{130325}{756} - \frac{796069}{2016} \eta + \frac{100019}{864} \eta^2 \right) \right .  \\
&& \left . + \eta \left(\frac{1195759}{18144} - \frac{257023}{1008} \eta + \frac{2903}{32} \eta^2 \right) \right ] \vec \chi_i \cdot \hat{\rm L}_{\rm N}
\end{eqnarray*}
where $q_i$ is a quadrupole-moment parameter, $\hat {\rm L}_{\rm N}$ is the unit vector in the direction of the binary's orbital angular momentum, and $\vec \chi_i = \vec S_i / m_i^2$ are the dimensionless spins of the $ith$ body.  In the works that derive these pN corrections, $q_i$ has been implicitly set to 1. This is the value it takes for spinning black holes, but not the value it takes for neutron stars and possibly other spinning exotica [see for example Eq. (8) of Ref.~\cite{1998PhRvD..57.5287P}]. However, we adopt the same simplification here since we will not be considering spinning systems outside of the Kerr class.

We reparameterize the component spins $\chi_i$ into an antisymmetric and a symmetric combination,
\begin{eqnarray}
\label{eq:chis}
\vec \chi_s &=& \frac{1}{2} \left (\vec \chi_1 +\vec  \chi_2\right)  \\
\label{eq:chia}
\vec \chi_a &=& \frac{1}{2} \left( \vec  \chi_1 - \vec \chi_2 \right) \ .
\end{eqnarray}
Recall that we restrict ourselves to spin-aligned (or antialigned), nonprecessing waveforms, which means $\vec \chi_a \cdot \hat L_{\rm N} = \pm |\vec \chi_a|$ and $\vec \chi_s \cdot \hat L_{\rm N} = \pm |\vec \chi_s|$.  The positive sign corresponds to systems with (anti)symmetric spins aligned with the orbital angular momentum of the binary, and the negative sign corresponds to systems with (anti)symmetric spins antialigned with the binary's orbital angular momentum.  

We also study nonspinning waveforms that include the 5.0 pN-order tidal correction to the phase.  Tidal corrections are calculated for the phase beyond 5.0 pN order \cite{tidalcorrections}.  However, we find that the tidal corrections beyond 5.0 pN order in phase are completely unmeasurable by the Fisher matrix.  Including these terms create a worse-conditioned Fisher matrix and does not affect the measurability of the 5.0 pN-order tidal correction.  Therefore, we omit all but the leading-order tidal correction in this work.  

The point-particle contributions to the phase of the waveform are only calculated through 3.5 pN order ($v_k^7$ beyond leading order).  The leading-order tidal correction to the phase appears at 5.0 pN order ($v_k^{10}$ beyond leading order).  Therefore, the 5.0 pN-order term in the phase of the waveform does not include point-particle effects.  The 5.0 pN-order tidal term that adds linearly to Eq.~\eqref{eq:spinphase} is
\begin{equation}
\label{eq:tidalcorrection}
\delta\Psi_{\rm tidal}(v_k) = -\frac{117 \tilde \Lambda}{16 \eta} v_k^5\ ,
\end{equation}
with $\tilde \Lambda = \tilde \lambda / M^5$ and
\begin{equation}
\label{eq:tildelambda}
\tilde \lambda = \frac{1}{26} \left( \frac{m_1 + 12 m_2}{m_1} \lambda_1 + \frac{m_2 + 12m_1}{m_2} \lambda_2 \right) \ ,
\end{equation}
where $\lambda_i$ is the tidal deformability parameter for component mass $m_i$ \cite{Ben}.  The tidal deformability parameter, which in this post-Newtonian description describes the ratio of the induced quadrupole moment to the perturbing external tidal field, is written in terms of the dimensionless tidal Love number $k_2$ \cite{Ben} as
\begin{equation}
\label{eq:tidallovenumber}
\lambda = \frac{2}{3} k_2 r^5 \ ,
\end{equation}
with $r$ being the radius of the star.  A fully relativistic generalization of this was provided in Ref.~\cite{Binnington}, where it was shown that for nonrotating black holes, the relativistic Love numbers all vanish. This remains true even when the black hole is deformed by a tidal field.

We examine two scenarios: spinning systems with no tidal corrections and nonspinning systems with tidal corrections.  For spinning systems we ``turn on" the 1.5 pN- to 2.5 pN-order spin corrections in the phase by setting the parameter $\epsilon=1$ in Eq.~\eqref{eq:spinphase}, and we ``turn on" the 3.0 pN- and 3.5 pN-order spin corrections in the phase by setting the parameter $\alpha = 1$.  We also turn on or off the spin corrections in the amplitude of the waveform as derived in Ref.~\cite{Evan}.  For nonspinning systems with tidal corrections, we turn off all of the spin corrections in the phase and the amplitude and add Eq.~\eqref{eq:tidalcorrection} linearly to Eq.~\eqref{eq:spinphase} for the phase of the waveform.  We do not include any tidal corrections in the amplitude of the waveform, because they have not yet been calculated.

\section{Parameter Estimation}
\label{sec:fishersvd}

\subsection{Fisher matrix}

We construct the covariance matrix using the Fisher information matrix formalism for a single detector to determine parameter errors and correlations.  For a large enough signal-to-noise ratio (SNR), the measurement errors on the  waveform parameters $\vec{\theta}$ given a gravitational waveform $\tilde h(f)$ fall into a Gaussian probability density function
\begin{equation*}
p(\Delta \vec{\theta}) = \sqrt{{\rm det} \left(\frac{\mathbf{\Gamma}}{2\pi}\right)} e^{\left(-\frac{1}{2}\Gamma_{ij} \Delta \theta^i \Delta \theta^j\right)} \ ,
\end{equation*}
where $\mathbf{\Gamma}$ is the Fisher information matrix \cite{Jolien, Cutler}.  The components of the Fisher matrix are defined as
\begin{equation}
\Gamma_{ij} = \left . \left(\frac{\partial h}{\partial \theta^i} \right | \left . \frac{\partial h}{\partial \theta^j}  \right) \right |_{ \vec\theta_{\rm max}} \ ,
\end{equation}
where $h$ is the gravitational waveform, $\theta^{i}$ is a waveform parameter, $\vec\theta_{\rm max}$ is the set of true parameters, and $( \cdot\cdot\cdot \mid \cdot\cdot\cdot )$ is an inner product defined by
\begin{equation}
\label{eq:ip}
\left( a \mid b \right) = 4 {\rm Re} \int_0^\infty \frac{\tilde a (f) \tilde b^* (f)}{S_n(f)} df 
\end{equation}
for power spectral density $S_n(f)$. 

The root-mean-square error on a parameter $\theta^{i}$ is derived from the inverse Fisher matrix, which is the covariance matrix under certain assumptions \cite{Michele},
\begin{equation}
\left(\Delta \theta^i\right)_{\rm rms} = \sqrt{\left(\Gamma^{-1}\right)_{ii}} \mbox{ (no summation over $i$)}\ .
\end{equation}
The correlation between two parameters $\theta^i$ and $\theta^j$ is also derived from the inverse Fisher matrix,
\begin{equation}
c_{ij} = \frac{\left(\Gamma^{-1}\right)^{ij}}{\sqrt{\left(\Gamma^{-1}\right)^{ii}\left(\Gamma^{-1}\right)^{jj}}} \mbox{ (no summation over $i$ or $j$)} \ .
\end{equation}

\subsection{Validity of the Fisher matrix}

\begin{figure*}[t]
	\begin{minipage}[b]{0.4\textwidth}
	\centering
	\includegraphics[width=\textwidth]{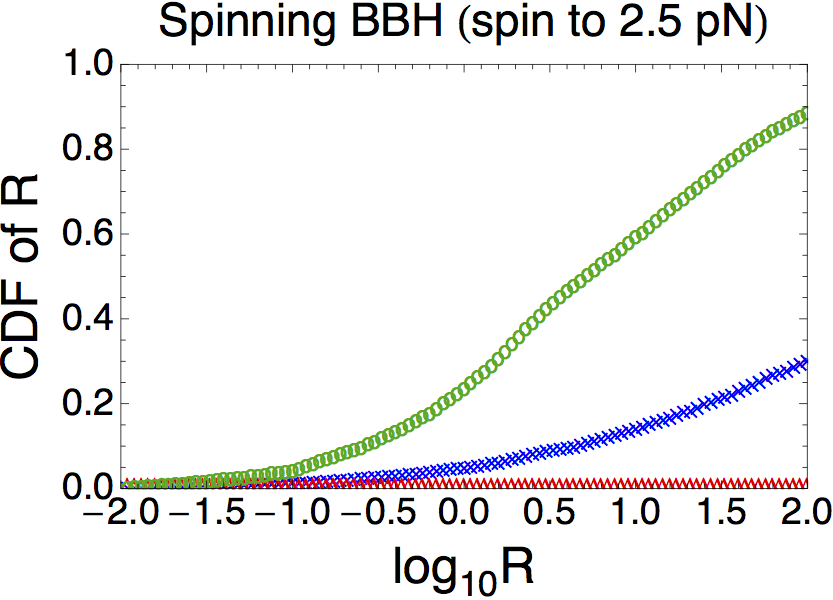}
	\end{minipage}
	~
	~
	\begin{minipage}[b]{0.4\textwidth}
	\centering
	\includegraphics[width=\textwidth]{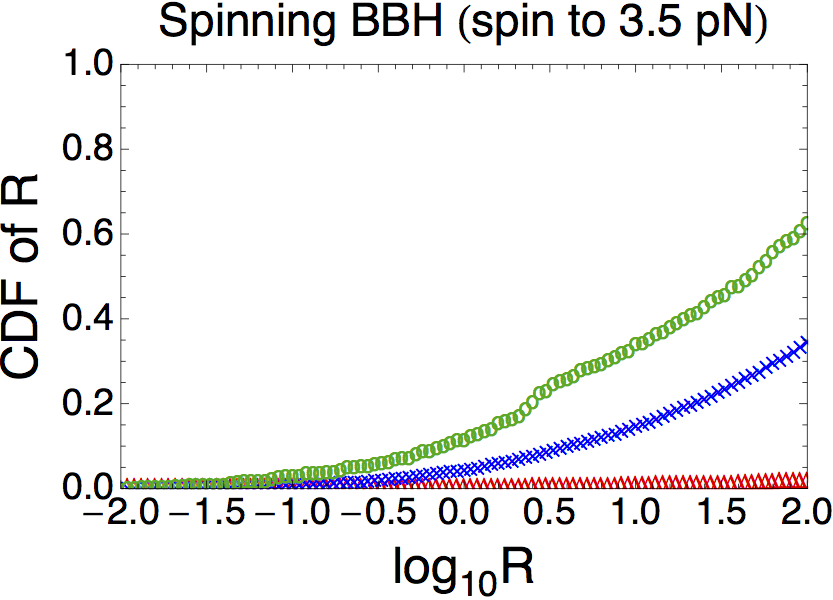}
	\end{minipage}
	
	\
	
	\begin{minipage}[b]{0.4\textwidth}
	\centering
	\includegraphics[width=\textwidth]{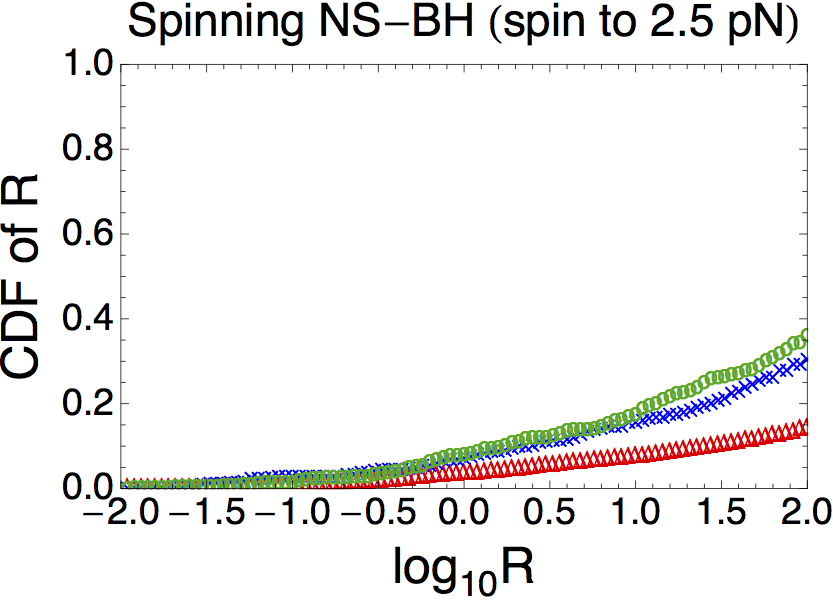}
	\end{minipage}
	~
	~
	\begin{minipage}[b]{0.4\textwidth}
	\centering
	\includegraphics[width=\textwidth]{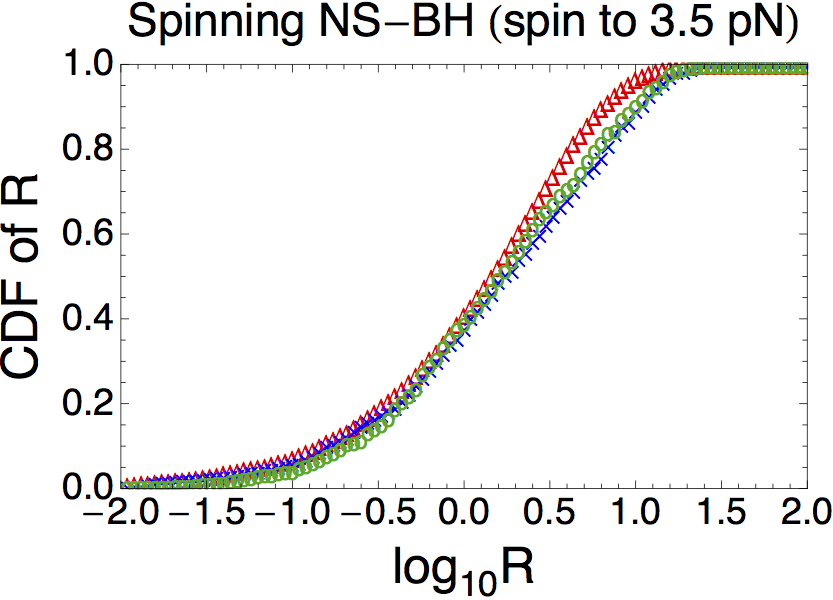}
	\end{minipage}
	
	\
	
	\begin{minipage}[b]{0.4\textwidth}
	\centering
	\includegraphics[width=\textwidth]{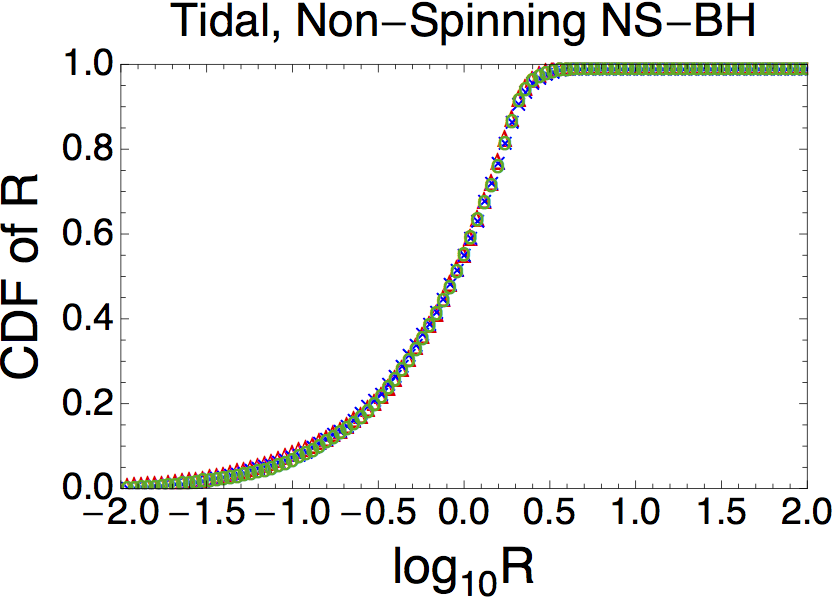}
	\end{minipage}
\caption{\footnotesize The cumulative distribution function (CDF) for the quantity $\log_{10} R$ where $R \equiv |\log r|$.  The quantity $R$, defined in Eq.~\eqref{eq:logr}, is a measure of the self-consistency of the Fisher matrix.  This quantity is calculated at 5000 random points on the $1\sigma$-error surface, and the CDF of these points is plotted here.  The smaller values of $R$ indicate a more self-consistent Fisher matrix.  Therefore, the most self-consistent Fisher matrix calculations have a CDF of $R$ that rises quickly.  Above we plot $\log_{10} R$ for a spinning BBH system ($m_1 = 10 \mbox{ M}_\odot$, $m_2 = 11 \mbox{ M}_\odot$), a spinning NS-BH system ($m_1 = 1.4 \mbox{ M}_\odot$, $m_2 = 10 \mbox{ M}_\odot$), and a tidal BBH system ($m_1 = 10 \mbox{ M}_\odot$, $m_2 = 11 \mbox{ M}_\odot$) all with $t_c = 0$, $\phi_c = 0$, $\theta = \pi/6$, $\phi = \pi/6$, $\psi = \pi/4$ and $\iota = \pi/3$ and with a fixed SNR of $\rho = 100$.  For the spinning BBH system, the component spins are $\chi_1 = \chi_2 = 1$, for the spinning NS-BH system the component spins are $\chi_1=0$ and $\chi_2=1$, and for the nonspinning, tidal BBH system the tidal deformability parameter is $\tilde \Lambda = 0$.  The $1\sigma$-error estimates employed in this calculation were obtained from a five- or four-parameter Fisher matrix calculation with $\vec \theta = \{\log \mathcal{M}, \eta, t_c, \chi_a, \chi_s\}$ or $\vec \theta = \{\log \mathcal{M}, \eta, t_c, \tilde \Lambda\}$ for the spinning systems and the tidal system, respectively.  The plot shows results for the Newtonian-amplitude waveform (red triangles), the 0.5 pN order amplitude-corrected waveform (blue X's), and the 1.0 pN order amplitude-corrected waveform (green circles), with spin corrections included in the amplitude for the spinning systems.  The spinning-system plot titles indicate which spin corrections are kept in the phase of the waveform.}
\label{fig:fishervalidity}
\end{figure*}

The Fisher matrix provides an approximation to the covariance matrix that represents the Cramer-Rao bound \cite{Michele}.  Studies using the Fisher matrix in the context of gravitational-wave parameter estimation are vast in the literature (e.g. Refs.~\cite{poissonandwill, Cutler, Finn1992, Alex,Ben, Chris}).  However, there are several drawbacks in employing the Fisher matrix for parameter estimation studies.  The derivation of the Fisher matrix requires the linearized signal approximation (LSA), which is only valid in the high-SNR limit \cite{Michele}.  Real gravitational-wave detections in the advanced-detector era are not expected to fall into the high-SNR limit \cite{rates}.  In addition, the Fisher matrix assumes a Gaussian, single-modal distribution of the likelihood function \cite{Michele, CarlFisher}.  In reality, the likelihood could be very non-Gaussian and multimodal.  The Fisher matrix does not fully explore the parameter space, but rather focuses on one point in parameter space and assumes a Gaussian likelihood about this point.  In reality, a full Bayesian calculation of the likelihood function starting from the raw data and using techniques such as MCMC to explore parameter space is required for accurate parameter estimation, which has also been studied extensively in the literature (e.g. Refs.~\cite{VanDerSluys2008, VanDerSluysetal2008, Veitch2010, Raymond2010, Veitch2012, PEPaper, CarlFisher}).  Rodriguez {\it et al.} \cite{CarlFisher} perform an in-depth comparison of the Fisher matrix with a full Bayesian MCMC study and find that the Fisher matrix can be very ill suited to parameter estimation for certain systems.  Below, we perform some tests to verify the validity of the Fisher matrix approach in our work.

Vallisneri discusses a self-consistency check for the Fisher matrix in Ref.~\cite{Michele}.  To determine the level of self-consistency of the Fisher matrix, we calculate
\begin{equation}
\label{eq:logr}
\left | \log r \right | = \frac{1}{2}\left(\left . (\Delta\theta^j)_{\rm rms} h_j - \Delta h\right |(\Delta\theta^k)_{\rm rms} h_k - \Delta h \right)
\end{equation}
where $h_j = \left . \partial h/\partial \theta_j \right |_{\vec \theta_{\rm max}}$, $\Delta h = h|_{\vec \theta_{1\sigma}} - h|_{\vec \theta_{\rm max}}$, and $\vec \theta_{1\sigma}$ is a point in parameter space that lies on the $1\sigma$-error surface.  The value of $|\log r |$ will depend on the SNR, since the $1\sigma$-error surface and parameter root-mean-square errors are a function of SNR.

Fig.~\ref{fig:fishervalidity} plots the cumulative distribution function of $\log_{10} R$ ($R \equiv |\log r|$) calculated for a large number of random points on the $1\sigma$-error surface at a fixed SNR of 100 for the Newtonian-amplitude waveform (red triangles), the 0.5 pN-order amplitude-corrected waveform (blue X's), and the 1.0 pN-order amplitude-corrected waveform (green circles), with spin corrections included in the amplitude for the spinning systems.  The $1\sigma$-error surface used in the calculation of $|\log r|$ came from a five- or four-parameter Fisher matrix calculation with $\vec \theta = \{\log \mathcal{M}, \eta, t_c, \chi_a, \chi_s\}$ or $\vec \theta = \{\log \mathcal{M}, \eta, t_c, \tilde \Lambda \}$ for the spinning systems and the nonspinning, tidal system, respectively.  Fig.~\ref{fig:fishervalidity} shows results for the spinning systems both with and without the 3.0 pN- and 3.5 pN-order spin corrections to the phase.

Fig.~\ref{fig:fishervalidity} indicates that the approximations necessary for the Fisher matrix formalism to be self-consistent, such as the linearized signal approximation (LSA), are more valid for the 1.0 pN-order amplitude-corrected waveform with spin corrections in the amplitude when compared to the Newtonian-amplitude waveform and the 0.5 pN-order amplitude-corrected waveform for the spinning systems.  In addition, including the 3.0 pN- and 3.5 pN-order spin corrections to the phase for the NS-BH system leads to significant improvement in the self-consistency of the Fisher matrix.  However, the spinning BBH system is either left unchanged or made slightly less valid by including these higher-order spin-orbit corrections.  For the nonspinning, tidal BBH system, all of the waveforms prove equally valid.  

Vallisneri notes in Ref.~\cite{Michele} that the LSA will be more valid for parameter spaces with weaker correlations.  As will be discussed in Sec.~\ref{sec:results}, the amplitude-corrected waveforms cause certain parameters that are strongly correlated in the Newtonian-amplitude waveform to decouple for the spinning BBH system.  Parameter correlations are broken when moving both from the Newtonian-amplitude waveform to the 0.5 p- order amplitude-corrected waveform and from the 0.5 pN-order amplitude-corrected waveform to the spin-dependent 1.0 pN-order amplitude-corrected waveform.  Degeneracies are also slightly decreased when including the 3.0 pN- and 3.5 pN-order spin corrections in the phase for the spinning NS-BH system but mostly unchanged for the spinning BBH system.

Fig.~\ref{fig:fishervalidity} is a good reference for the self-consistency of the Fisher matrix for different orders of the post-Newtonian expansion.  The scale of $| \log r|$ indicates that the Fisher matrix may only be self-consistent for high SNR.  Therefore, we perform an additional investigation into the validity of the Fisher matrix below.  The results of this investigation conclude that the Fisher matrix should give fairly reliable results for the cases studied in this work, even for a SNR of 10.  

\begin{figure*}[t]
	\begin{minipage}[b]{0.4\textwidth}
	\centering
	\includegraphics[width=\textwidth]{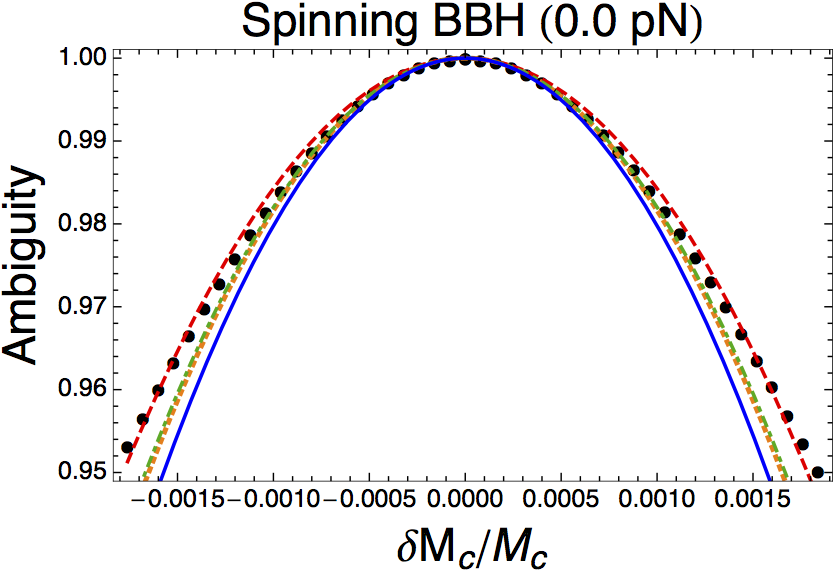}
	\end{minipage}
	~
	~
	\begin{minipage}[b]{0.4\textwidth}
	\centering
	\includegraphics[width=\textwidth]{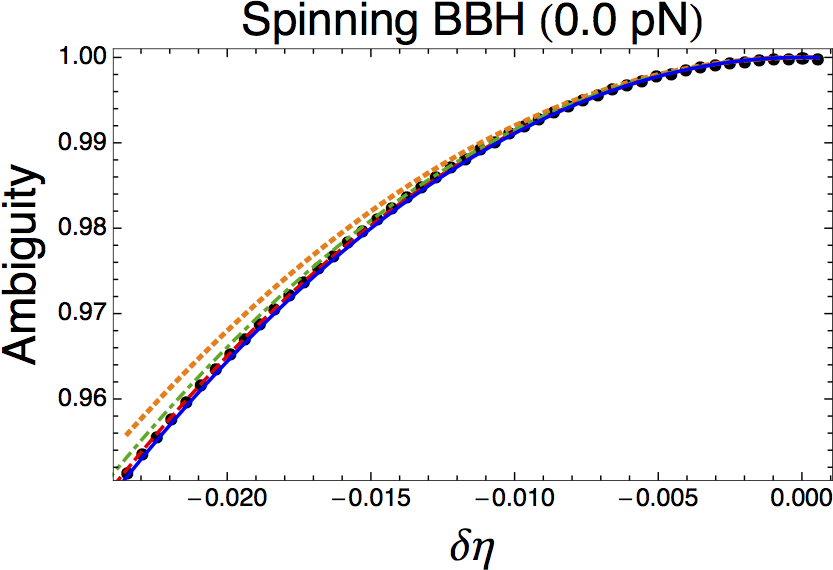}
	\end{minipage}
	
	\
	
	\begin{minipage}[b]{0.4\textwidth}
	\centering
	\includegraphics[width=\textwidth]{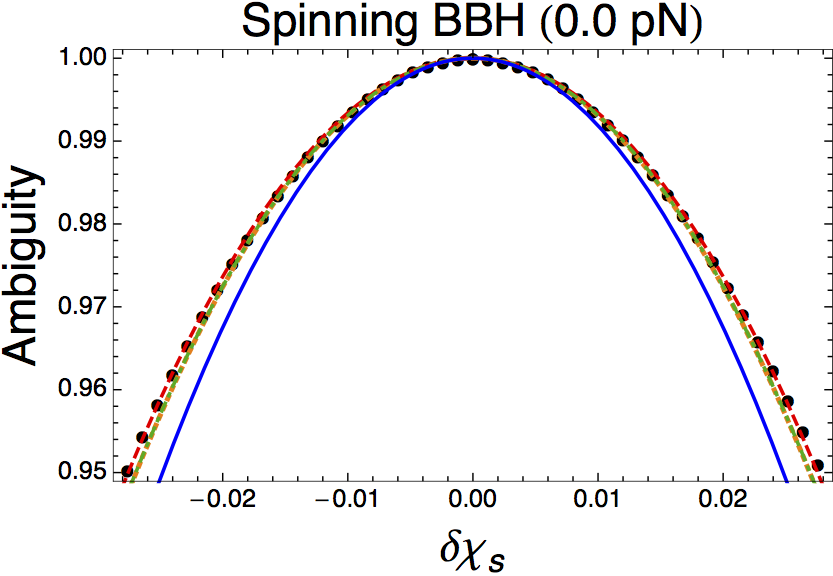}
	\end{minipage}
	~
	~
	\begin{minipage}[b]{0.4\textwidth}
	\centering
	\includegraphics[width=\textwidth]{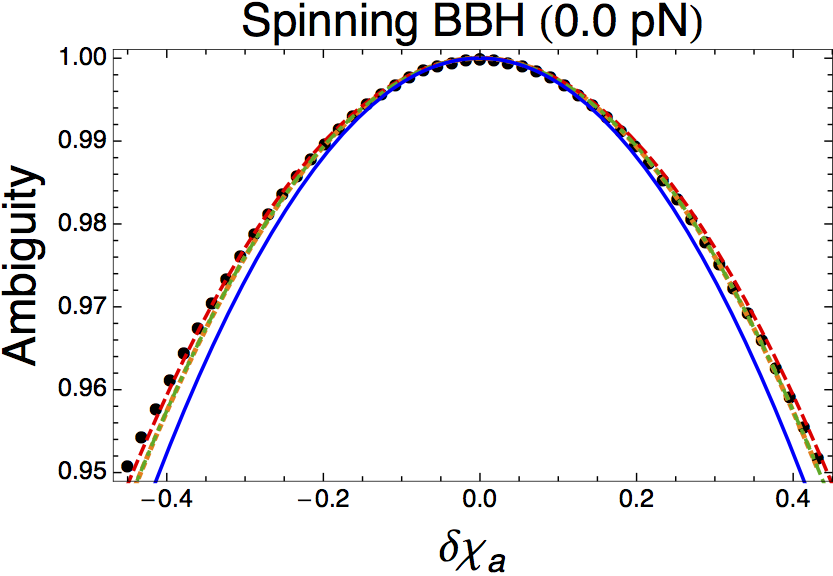}
	\end{minipage}
	
	\caption{The ambiguity function for different systems, as indicated by the title of each plot, over the most relevant parameters to this work, $\left\{{\mathcal M}, \eta, \chi_s, \chi_a \right\}$.  The pN order in each title references the pN expansion order of the amplitude of each waveform.  In each plot, the parameter on the $x$ axis is varied while all other parameters are held fixed at fiducial values (see Sec.~\ref{sec:params}).  Each plot also shows quadratic fits over three different scales: $P\ge0.95$ (red dashed line), $P \ge 0.99$ (green dot-dashed line), and $P \ge 0.999$ (blue solid line).  The actual ambiguity function is shown with black dots.  The fit lines are all fairly close to each other, which indicates that the likelihood for these systems is fairly Gaussian over the relevant scale.  In addition, the orange, dotted line shows the quadratic fit predicted from the Fisher matrix, which is also in good agreement.}
	\label{fig:ambiguity}
\end{figure*}

The Fisher matrix involves the partial derivative of the waveform with respect to a set of parameters.  In order for the Fisher matrix approximation to be valid, the likelihood needs to be fairly Gaussian on scales appropriate to the SNR being studied.  One way to examine the Gaussianity of the likelihood would be through the ambiguity function $P$, defined in Ref.~\cite{Cho2013} as
\begin{equation}
P(\vec \theta_{\rm max},\vec \theta) = {\rm max}_{t_c, \psi} \frac{\left (\left .h_{\vec\theta_{\rm max}}\right |h_{\vec \theta}\right)}{\sqrt{\left(\left . h_{\vec\theta_{\rm max}}\right|h_{\vec\theta_{\rm \max}}\right)\left(\left . h_{\vec \theta}\right |h_{\vec\theta}\right)}}
\end{equation}
where ${\rm max}_{t_c, \psi}$ means a maximization over coalescence time and polarization angle, as described in Ref.~\cite{Cho2013}.  The ambiguity function is a measure of the overlap between the true waveform with parameters $\vec \theta_{\rm max}$ and a waveform described by parameters $\vec \theta$.

If the likelihood is Gaussian, the ambiguity function should fit well to a quadratic curve \cite{Cho2013}.  The scale over which the ambiguity function should be quadratic is determined by the SNR.  For a SNR of $\rho$, the ambiguity function should be well fit to the same Gaussian over scales up to $P \ge 1 - 1/\rho^2$ \cite{Cho2013}.  Throughout this work, we mainly study a SNR of $\rho = 10$, so the scale of interest for the ambiguity function is $P \ge 0.99$.  For completeness, we examine the ambiguity function on scales $P \ge 0.95$.  Fig.~\ref{fig:ambiguity} shows the ambiguity function over the most relevant parameters to this work, $\left\{{\mathcal M}, \eta, \chi_s, \chi_a \right\}$.  In each plot, the parameter on the $x$ axis is varied while all other parameters are held fixed at fiducial values, which are outlined for the previous validity test.  Each plot also shows quadratic fits over three different scales: $P\ge0.95$ (red dashed line), $P \ge 0.99$ (green dot-dashed line), and $P \ge 0.999$ (blue solid line).  The actual ambiguity function is shown with black dots.  Although we only show plots for the spinning BBH system in the Newtonian-amplitude, the plots look very similar for the different systems studied in this work and across different post-Newtonian approximations to the amplitude and phase.  The quadratic fits across different scales match up well.  This test indicates that the likelihood is appropriately Gaussian for the SNR studied in this work.

Fig.~\ref{fig:ambiguity} also shows the quadratic fit as predicted by the Fisher matrix (orange dotted line).  The comparison between ambiguity and the Fisher matrix is most simply seen by examining the logarithm of the Gaussian likelihood, as retrieved from Eqs. (17) and (22) in Ref.~\cite{Cho2013}, for example.  The one-dimensional ambiguity function over parameter $\theta_i$ not maximized over $t_c$ or $\psi$, denoted below as $\tilde P$, is simply related to the relevant Fisher matrix component $\Gamma_{ii}$,
\begin{equation}
\label{eq:ambfisher}
\tilde P = 1 - \frac{1}{2} \frac{\Gamma_{ii}}{\rho^2} (\Delta \theta_i)^2 \ \ \mbox{(no summation over $i$).}
\end{equation}
However, to make a more direct comparison with the normalized ambiguity function maximized over $t_c$ and $\psi$, the parameters $D_{\rm M}$, $t_c$, and $\psi$ should be projected out of the Fisher matrix.  Projecting out these three parameters is achieved by computing a four-dimensional Fisher matrix including the parameters of interest, $D_{\rm M}$, $t_c$, and $\psi$, inverting this matrix, and taking the inverse of the relevant component $\left[(\Gamma^{-1})_{ii}\right]^{-1}$.  The orange dotted lines plotted in Fig.~\ref{fig:ambiguity} are for the quadratic fit,
\begin{equation*}
P = 1 - \frac{1}{2} \frac{\left[(\Gamma^{-1})_{ii}\right]^{-1}}{\rho^2} (\Delta \theta_i)^2 \ \ \mbox{(no summation over $i$),}
\end{equation*}
where $\theta_i$ is either $\mathcal{M}$, $\eta$, $\chi_s$, or $\chi_a$.  These fits are very consistent with the ambiguity function calculation in all cases.

Qualitatively, we expect the Fisher matrix results to be accurate.  Quantitatively, the Fisher matrix results will be most accurate for a high SNR.  The results in this paper are provided for a SNR of 10.  The Fisher matrix results scale very simply from a SNR of 10 if the reader wishes to study different SNR scenarios.  Other sources of quantitative error that may exceed the errors introduced by the Fisher matrix are errors associated with the inaccuracies of the post-Newtonian waveforms.  When working with real data, additional quantitative errors, such as calibration errors, can also become significant.  This work is intended to give insight into the ability of aLIGO to study tests of general relativity in a mainly qualitative manner.  This study should motivate full Bayesian studies that will be required to investigate low-SNR scenarios quantitatively.  

\subsection{Singular-value decomposition}

The parameter spaces that we investigate can be 11 or 10 dimensional; see Eqs.~\eqref{eq:spinparamsfull} and~\eqref{eq:tidalparams}.  In these multidimensional parameter spaces, the Fisher matrix is often singular or badly conditioned and therefore difficult to invert.  One way we address this is by using a singular-value decomposition (SVD) on the Fisher matrix \cite{FisherSVD}.  The SVD of a matrix $\mathbf{\Gamma}$ is
\begin{equation}
\mathbf{\Gamma} = \mathbf{U} \mathbf{S} \mathbf{V^\dagger} \ ,
\end{equation}
where $\mathbf{S}$ is a diagonal matrix whose diagonal elements contain the singular values, and $\mathbf{U}$ and $\mathbf{V}$ are unitary matrices of the left and right singular vectors, respectively.  The covariance matrix in terms of its singular-value decomposition is
\begin{equation*}
\mathbf{\Gamma^{-1}} = \mathbf{V}\mathbf{S^{-1}} \mathbf{U^{\dagger}} \ .
\end{equation*}
Since the Fisher matrix is real and symmetric by definition, for our
case we have $\mathbf{V}=\mathbf{U}$ and this matrix will be an orthogonal matrix of the
real eigenvectors of $\mathbf{\Gamma}$.  

If the Fisher matrix is singular or badly conditioned, its singular values will be zero or very small.  We remove the singular or badly conditioned pieces of the Fisher matrix by zeroing out the elements of $\mathbf{S^{-1}}$ that are very large or infinite.  These elements correspond to the zero or very small singular values of the Fisher matrix, which become infinite or very large upon inversion.  Zeroing out these elements is effectively removing the unmeasurable linear combinations of parameters from the Fisher matrix.  In this way we are able to obtain error estimates for only the measurable parameters, and we do not have to assume {\it a priori} which are the measurable parameters.

\section{Parameters and Parameter-Space Bounds}
\label{sec:params}

\subsection{Spinning waveform}
\label{sec:spinparams}

For the spinning waveform described in Sec.~\ref{sec:waveforms}, the full parameter space is 11 dimensional,
\begin{eqnarray}
\nonumber
\vec \theta_{\rm spin, full} &=& \left\{\log (1/D_M), \log \mathcal{M}, \eta, t_c, \phi_c, \right .\\
\label{eq:spinparamsfull}
&& \left . \cos\iota,\chi_a, \chi_s, \cos\theta, \phi, \psi\right\}
\end{eqnarray}
where $\iota$ is the inclination angle of the binary, $\theta$ and $\phi$ are the sky position polar coordinates, $\psi$ is the polarization angle, and $\chi_s$ and $\chi_a$ are the symmetric and antisymmetric spin parameters described in Sec.~\ref{sec:waveforms}.  We use true values of $t_c=0$, $\phi_c=0$, $\iota=\pi/3$, $\theta=\pi/6$, $\phi=\pi/6$, and $\psi=\pi/4$ for all of the results reported here.  All calculations are performed for a fixed SNR, which determines the value of $D_M$ for each calculation.  The component masses and spins are varied as described in Sec.~\ref{sec:spinresults}.

We find that a smaller dimensional parameter space is required to obtain reliable results from the Fisher matrix when performing calculations with the Newtonian-amplitude spinning waveform, even when employing the SVD method described in Sec.~\ref{sec:fishersvd}.  For the Newtonian-amplitude spinning waveform calculations, we use a reduced six-dimensional parameter space:
\begin{equation}
\label{eq:spinparamsred}
\vec \theta_{\rm spin, reduced} = \left\{\log (1/D_M), \log \mathcal{M}, \eta, t_c, \chi_a, \chi_s\right\} \ .
\end{equation}
For this reduced parameter space, we use true values of $t_c=0$, $\phi_c=0$, $\iota=\pi/3$, $\theta=\pi/6$, $\phi=\pi/6$, and $\psi=\pi/4$, and we vary component masses and spins as described in Sec.~\ref{sec:spinresults}.  Once again, the fixed SNR for each calculation determines the value of $D_M$ for that system.  

We exploit bounds on the symmetric mass ratio and the Kerr parameter to reduce the acceptable parameter space.  The physical bounds on $m_1$ and $m_2$ and the definition of the symmetric mass ratio restrict $\eta$ to be $(0, 1/4]$.  For Kerr solutions, cosmic censorship requires $| \vec \chi_i | \leq 1$, which restricts $|\vec \chi_s|$ and $|\vec \chi_a|$ to be less than or equal to 1.  The bounds on spin and the symmetric mass ratio create a finite region of the two-dimensional spin-mass parameter space that is both physical and consistent with a Kerr black hole.  Excluding the unphysical areas of $\eta$ parameter space is not imposed as a flat prior in the Fisher matrix calculation but is applied after the fact to the error ellipse that results from an unrestricted Fisher matrix calculation.  A more detailed discussion on the improved measurability of spin by restricting the spin-mass parameter space can be found in Sec.~\ref{sec:spinresults}.  

For amplitude-corrected waveforms, the 11-dimensional parameter space given in Eq.~\eqref{eq:spinparamsfull} often leads to a badly conditioned or singular Fisher matrix.  We use the singular-value decomposition method discussed in Sec.~\ref{sec:fishersvd} to invert the Fisher matrix and discover the unmeasurable linear combinations of parameters.  For the Newtonian-amplitude waveform all of the parameters in the reduced parameter space $\vec \theta_{\rm spin, reduced}$ are measurable.  For the lowest-order amplitude-corrected waveform (0.5 pN), the measurable parameters are $\mathcal{M}$, $\eta$, $t_c$, $\phi_c$, $\cos\iota$, $\chi_a$, and $\chi_s$.  For the 1.0 pN order amplitude-corrected waveform, the measurable parameters are $\mathcal{M}$, $\eta$, $t_c$, $\phi_c$, $\cos\iota$, $\chi_a$, $\chi_s$, and $\phi$.  In Sec.~\ref{sec:spinresults} we only report on the measurement errors for $\mathcal{M}$, $\eta$, $\chi_s$ and $\chi_a$, since these are the most pertinent to our study.

\subsection{Nonspinning, tidal waveform}
\label{sec:tidalparams}

For the nonspinning, tidal waveform described in Sec.~\ref{sec:waveforms}, we investigate a 10--dimensional parameter space,
\begin{eqnarray}
\nonumber
\vec \theta_{\rm tidal} &=& \left\{\log (1/D_M), \log \mathcal{M}, \eta, t_c, \phi_c, \right . \\
\label{eq:tidalparams}
&&\left . \cos\iota, \tilde \Lambda, \cos\theta, \phi, \psi\right\} \ .
\end{eqnarray}
We use true values of $t_c=0$, $\phi_c=0$, $\iota=\pi/3$, $\theta=\pi/6$, $\phi=\pi/6$, and $\psi=\pi/4$ for all of the results reported here.  All calculations are performed for a fixed SNR, which determines the value of $D_M$ for each calculation.  The component masses and the tidal parameter are varied as described in Sec.~\ref{sec:tidalresults}.

As was the case with the spinning waveform, the tidal parameter space also has bounds with useful physical interpretations.  We explore how exploiting the physical bound on the symmetric mass ratio ($0 < \eta \le 1/4$) affects the measurability of the tidal parameter.  In addition, we place a bound on the tidal deformability parameter ($\tilde \Lambda = 0$) for the waveform to be consistent with expectations from the no-hair theorem, in the sense described in Sec.~\ref{sec:intro}.  Previous work on tidal deformability calculations for compact systems \cite{Nagar} suggests that $\delta \Psi_{\rm tidal}$ should be zero or small for black holes.  The closest matter analog would be an incompressible star at maximum compactness ($c = m/r = 4/9$), for which the tidal Love number would be $k_2 = 0.0017103$ \cite{Nagar}.  For an equal mass, equal radius binary system, the parameter $\tilde \Lambda$ is
\begin{equation*}
\tilde \Lambda = \frac{\tilde \lambda}{(2m)^5} = \frac{\lambda}{(2m)^5} = \frac{1}{48} k_2 \left(\frac{r}{m}\right)^5
\end{equation*}
where the above follows from the definition of $\tilde \lambda$ (given by Eq.~\eqref{eq:tildelambda}) for an equal mass system, $\lambda$ is the tidal parameter for one component object as defined in Eq.~\eqref{eq:tidallovenumber}, $r$ is the radius of one component object, and $m$ is the mass of one component object.  Using the ratio of $m/r$ for maximum compactness in the above expression gives $\tilde \Lambda \approx 0.002$.  Therefore, it is reasonable to conclude that the parameter $\tilde \Lambda$ should be small, if not identically zero, for black holes.  There could potentially be internal structure effects appearing at 5.0 pN order in the phase that differ from the point particle approximation, but these effects should be undetectable for a black hole to have no hair.   Therefore, we take $\delta \Psi_{\rm tidal} = 0$, which implies $\tilde \Lambda = 0$,  for a nonspinning black hole with no hair.  For the most comprehensive aLIGO test of the no-hair theorem, it would be more appropriate to use numerical relativity waveforms with various realizations of internal structure parameterized by $\tilde \Lambda$.

Just as with the spinning waveform, this 10--dimensional parameter space often leads to a badly conditioned or singular Fisher matrix.  Using the method described in Sec.~\ref{sec:fishersvd}, we determine the measurable parameters for each waveform.  For the Newtonian-amplitude waveform, the measurable parameters are $\mathcal{M}$, $\eta$, $t_c$, and $\tilde \Lambda$.  For the lowest-order amplitude-corrected waveform (0.5 pN), the measurable parameters are $\mathcal{M}$, $\eta$, $t_c$, $\phi_c$, $\cos\iota$, and $\tilde \Lambda$.  For the 1.0 pN order amplitude-corrected waveforms, $\phi$ also becomes measurable. In Sec.~\ref{sec:tidalresults} we only report on the measurement errors for $\mathcal{M}$, $\eta$, and $\tilde \Lambda$, since these are the most pertinent to our study.

\section{Results}
\label{sec:results}

\begin{figure*}[t]
	\begin{minipage}[b]{0.4\textwidth}
	\centering
		\includegraphics[width=\textwidth]{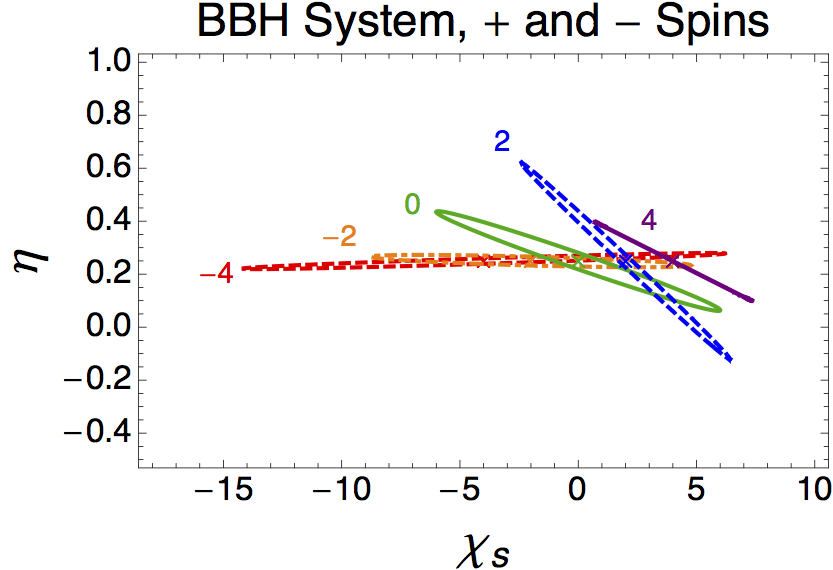}
	\end{minipage}
	~
	~	
	\begin{minipage}[b]{0.4\textwidth}
	\centering
		\includegraphics[width=\textwidth]{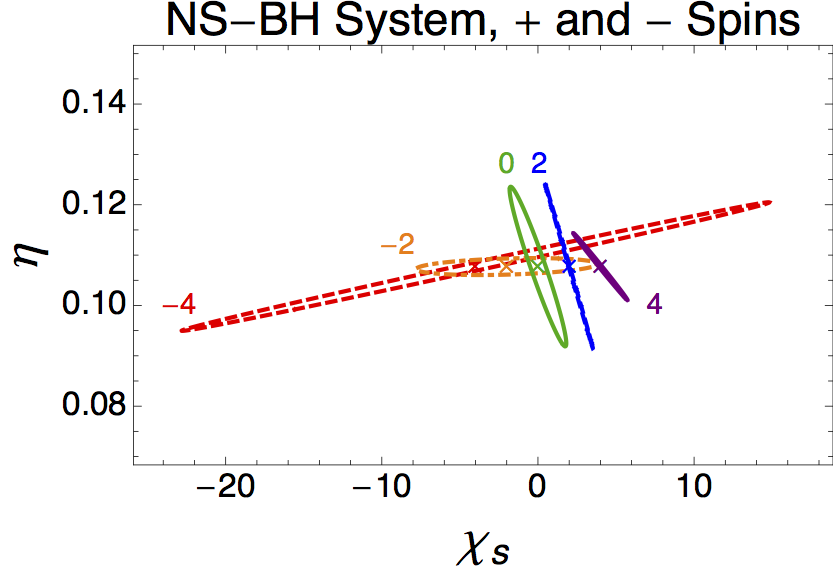}
	\end{minipage}
	\caption{\footnotesize These plots show 1$\sigma$ error ellipses in the $\eta-\chi_s$ parameter space for a spinning Newtonian-amplitude waveform with spin corrections in the phase to 2.5 pN order as described in Sec.~\ref{sec:waveforms} and with the reduced parameter space $\vec \theta_{\rm spin, reduced}$ described in Sec.~\ref{sec:spinparams}.  These ellipses are calculated for a spinning BBH system with $m_1 = 10 \mbox{ M}_\odot$ and $m_2 = 11 \mbox{ M}_\odot$ (left plot) and a spinning NS-BH system with $m_1 = 1.4 \mbox{ M}_\odot$ and $m_2 = 10 \mbox{ M}_\odot$ (right plot). Both systems have true parameters $t_c = 0$, $\phi_c = 0$, $\theta = \pi/6$, $\phi = \pi/6$, $\psi = \pi/4$ and $\iota = \pi/3$ and a fixed SNR of $\rho = 10$.  For the BBH system, the component spins are varied from $\chi_1 = \chi_2 = -4$ (red dashed ellipse) to $\chi_1 = \chi_2 = 4$ (purple solid ellipse).  Each ellipse takes a step of 2 in component spins.  For the NS-BH system, the component spins are varied from $\chi_1 = 0$, $\chi_2 = -8$ (red dashed ellipse) to $\chi_1 = 0$, $\chi_2 = 8$ (purple, solid ellipse).  Each ellipse takes a step of 2 in $\chi_s$, which corresponds to the black hole taking a step of 4 in its component spin.  The neutron star spin is held fixed at zero.  The numbers near each ellipse indicate the $\chi_s$ value for that ellipse (color coded).}
		\label{fig:spinacrossspace}
\end{figure*}

\begin{figure*}[t]
	\begin{minipage}[b]{0.4\textwidth}
	\centering
		\includegraphics[width = \textwidth]{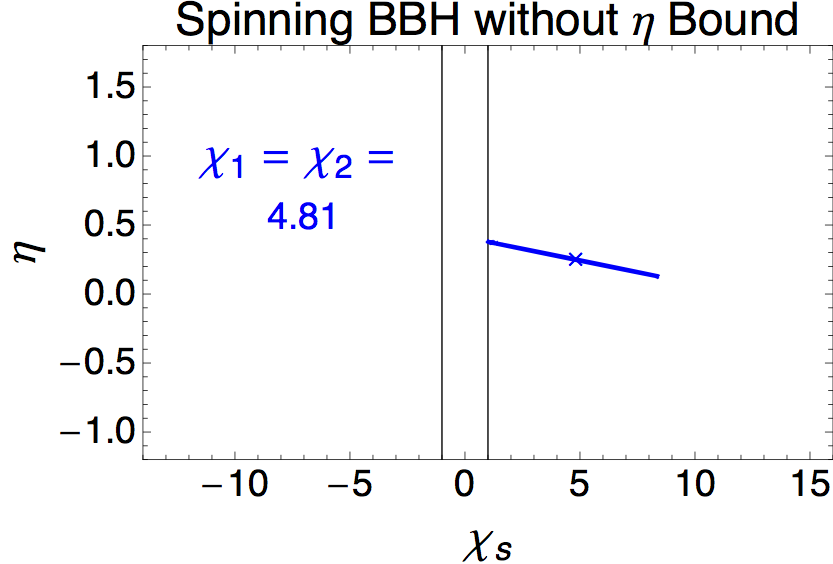}
	\end{minipage}
	~
	~
	\begin{minipage}[b]{0.4\textwidth}
	\centering
		\includegraphics[width = \textwidth]{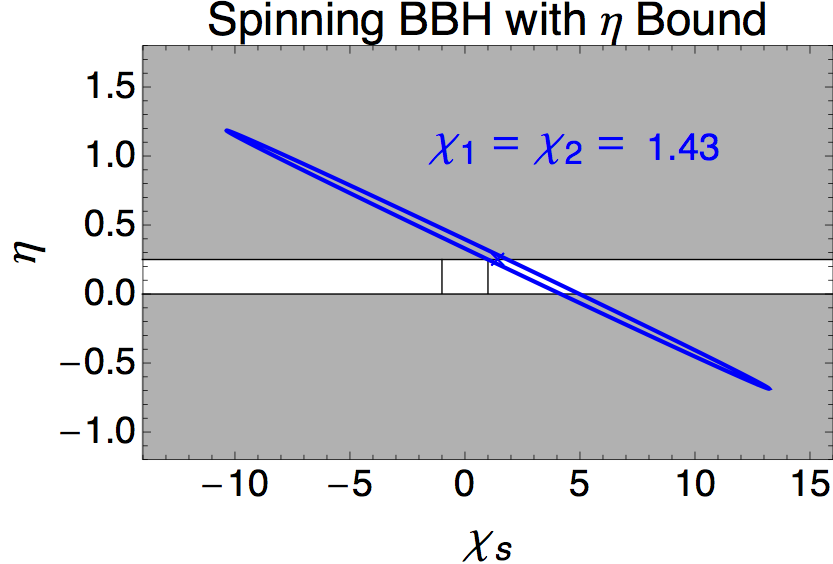}
	\end{minipage}
	\caption{\footnotesize These plots show 1$\sigma$ error ellipses in the $\eta-\chi_s$ parameter space for a spinning Newtonian-amplitude waveform with spin corrections in the phase to 2.5 pN order as described in Sec.~\ref{sec:waveforms} and with the reduced parameter space $\vec \theta_{\rm spin, reduced}$ described in Sec.~\ref{sec:spinparams}.  These ellipses are calculated for a spinning BBH system with $m_1 = 10 \mbox{ M}_\odot$, $m_2 = 11 \mbox{ M}_\odot$, $t_c = 0$, $\phi_c = 0$, $\theta = \pi/6$, $\phi = \pi/6$, $\psi = \pi/4$ and $\iota = \pi/3$ and with a fixed SNR of $\rho = 10$.  The component spins for each ellipse are given as an inlay on the plot.  The plot on the left shows the minimum detectable violation of the Kerr bound when considering the entire parameter space.  The plot on the right shows the minimum detectable violation of the Kerr bound when only considering the parts of the error ellipse that are physical.  The unphysical areas of parameter space are shaded gray in the plot on the right.  The vertical solid lines bound the region of parameter space that is consistent with the Kerr bound ($-1 \le \chi_s \le 1$).}
	\label{fig:spinetabound}
\end{figure*}

\subsection{Detectable apparent violations of the cosmic censorship conjecture}
\label{sec:spinresults}

We study two different spinning systems: a near equal mass binary black hole (BBH) system with component masses $m_1 = 10$ ${\rm M}_\odot$ and $m_2 = 11$ ${\rm M}_\odot$ and a neutron-star--black-hole (NS-BH) system with component masses $m_1 = 1.4$ ${\rm M}_\odot$ and $m_2 = 10$ ${\rm M}_\odot$.  The exactly equal mass limit is avoided due to singularities in the Fisher matrix at the equal mass limit when including amplitude corrections.  Both systems are parameterized as described by Eq.s~\eqref{eq:spinparamsfull} or~\eqref{eq:spinparamsred} and are subject to the parameter space bounds discussed in Sec.~\ref{sec:spinparams}.  We use the spinning waveform described in Sec.~\ref{sec:waveforms} with the phase kept to 3.5 pN order and the amplitude varied from 0.0 pN to 2.5 pN order.  Spin corrections are always included in the phase to 2.5 pN order.  We study the effect of turning on or off spin corrections in the amplitude of the waveform and turning on or off the 3.0 pN and 3.5 pN order spin corrections in the phase.  

We use the zero detuning, high power aLIGO power spectrum as given in \cite{PSD} for the power spectral density $S_n(f)$.   The inner product integrations are carried out from $f_{\rm min} = 10$ Hz to $f_{\rm max} = k F_{\rm LSO}$ where \cite{Jolien}
\begin{equation}
\label{eq:flso}
F_{\rm LSO} = \frac{1}{6^{3/2} 2 \pi M} \ .
\end{equation}

We choose to only examine positive (aligned) spins when determining the minimum detectable violation of the Kerr spin bound.  Negative (anti-aligned) spins are not as well measured as positive spins, and therefore will lead to a larger minimum detectable violation of the Kerr bound.  Fig.~\ref{fig:spinacrossspace} shows the $1\sigma$ error ellipses as produced by the Fisher matrix for both the spinning BBH system and the spinning NS-BH system.  Each ellipse is calculated for different values of component spin.  Fig.~\ref{fig:spinacrossspace} demonstrates how positive spins are more measurable than negative spins and therefore more useful in determining the minimum detectable violation.  The figure also illustrates how parameter measurability varies significantly for different values of spin for the BBH system and the NS-BH system.

One goal of our work with spinning black hole systems is to investigate how much better aLIGO would be able to detect a violation of the Kerr bound ($\chi_i > 1$) when only the physical area of $\eta$ parameter space is considered (see Sec.~\ref{sec:spinparams} for a discussion of parameter space bounds).  As mentioned before, this is not done by imposing a flat prior on the Fisher matrix.  Rather, an unrestricted Fisher matrix calculation is performed.  We examine the 1$\sigma$ error ellipses in the $\eta-\chi_s$ or $\eta-\chi_a$ plane and determine if the entire physical area of the ellipse is consistent or inconsistent with the Kerr bound.  We explore the parameter space until we find the minimum $\chi_i=j_i/(m_i)^2$ that violate the Kerr bound when considering only physical parts of the error ellipse.  As can be seen in Tables~\ref{tab:bbhspintable}--~\ref{tab:newspintermstableNSBH}, the parameter $\chi_s$ is better measured than the parameter $\chi_a$.  As discussed above and shown in Fig.~\ref{fig:spinacrossspace}, positive spins are also better measured than negative spins.  Therefore, we determine the minimum violation of the $\chi_s = 1$ bound in order to determine the minimum violation of the Kerr bound.

\begin{figure*}[p]
	\begin{minipage}[b]{0.38\textwidth}
		\centering
		\includegraphics[width = \textwidth]{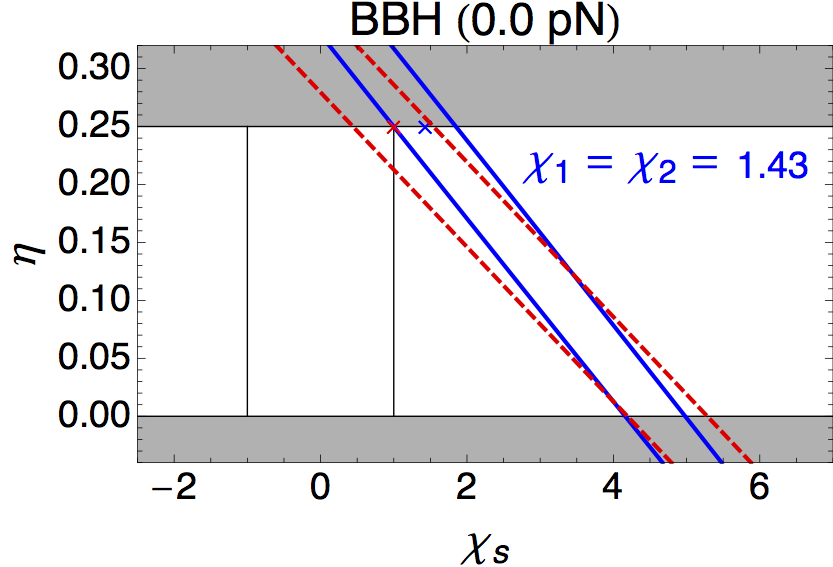}
	\end{minipage}
	~
	~
	\begin{minipage}[b]{0.38\textwidth}
	\centering
		\includegraphics[width = \textwidth]{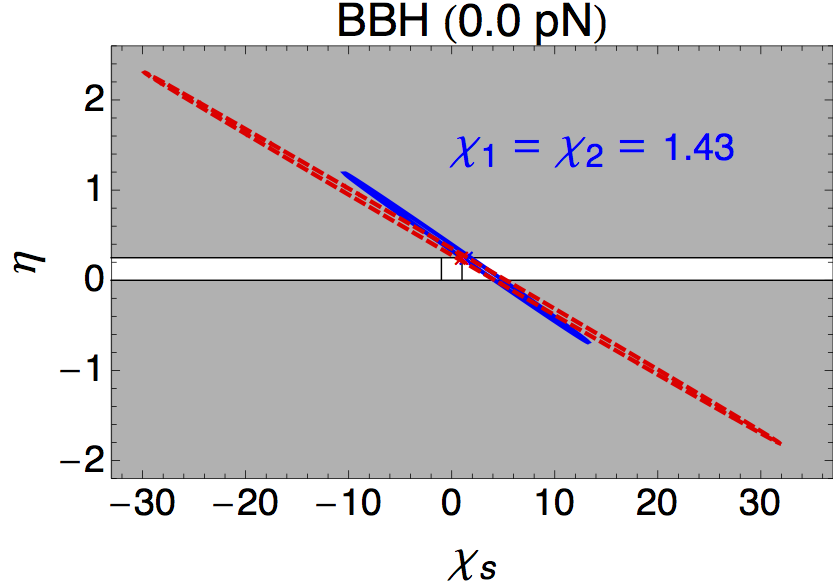}
	\end{minipage}
	
	\
	
	\begin{minipage}[b]{0.38\textwidth}
	\centering
		\includegraphics[width = \textwidth]{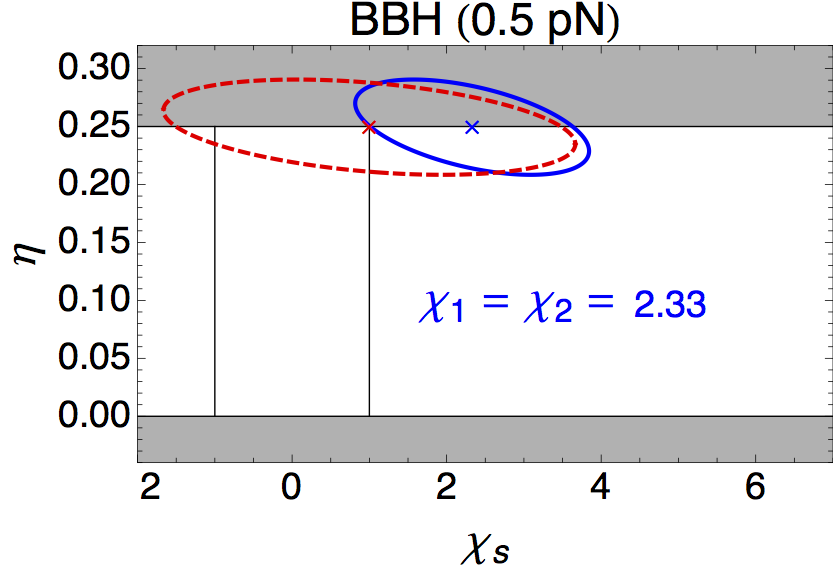}
	\end{minipage}
	~
	~
	\begin{minipage}[b]{0.38\textwidth}
	\centering
		\includegraphics[width = \textwidth]{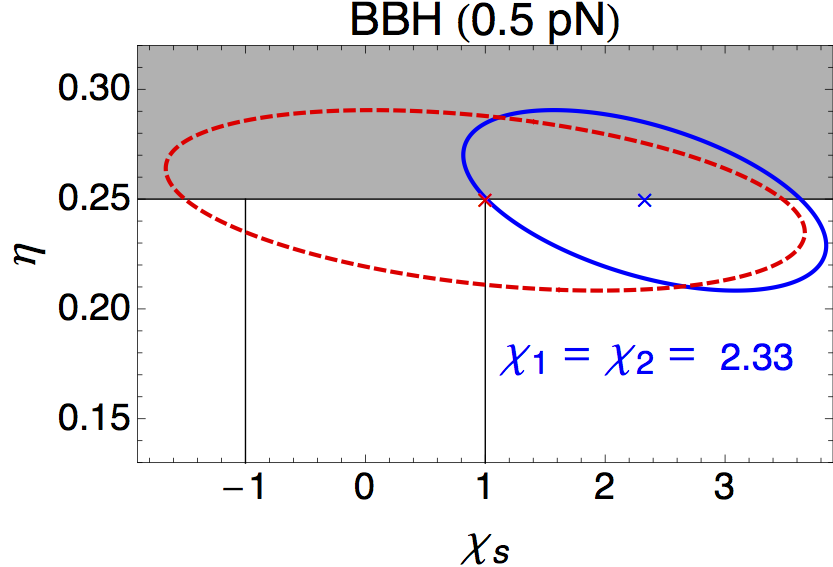}
	\end{minipage}
	
	\
	
	\begin{minipage}[b]{0.38\textwidth}
	\centering
		\includegraphics[width = \textwidth]{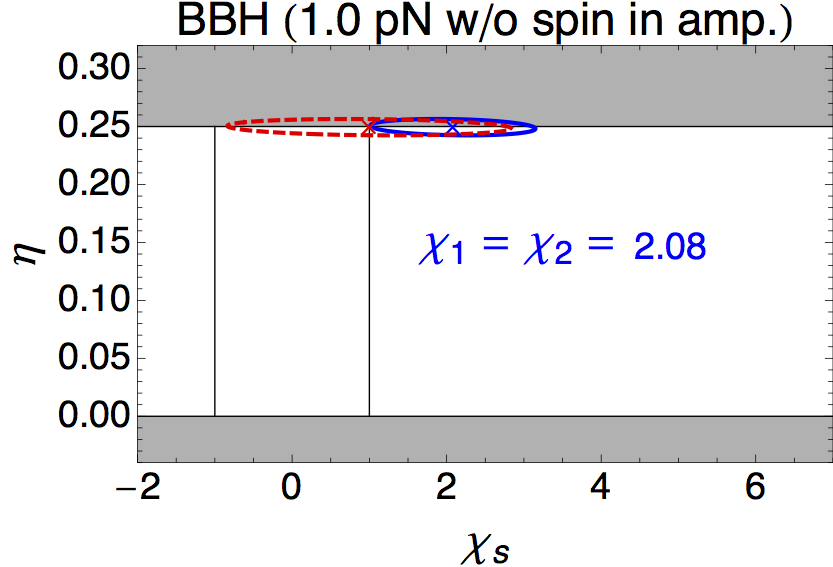}
	\end{minipage}
	~
	~
	\begin{minipage}[b]{0.38\textwidth}
	\centering
		\includegraphics[width = \textwidth]{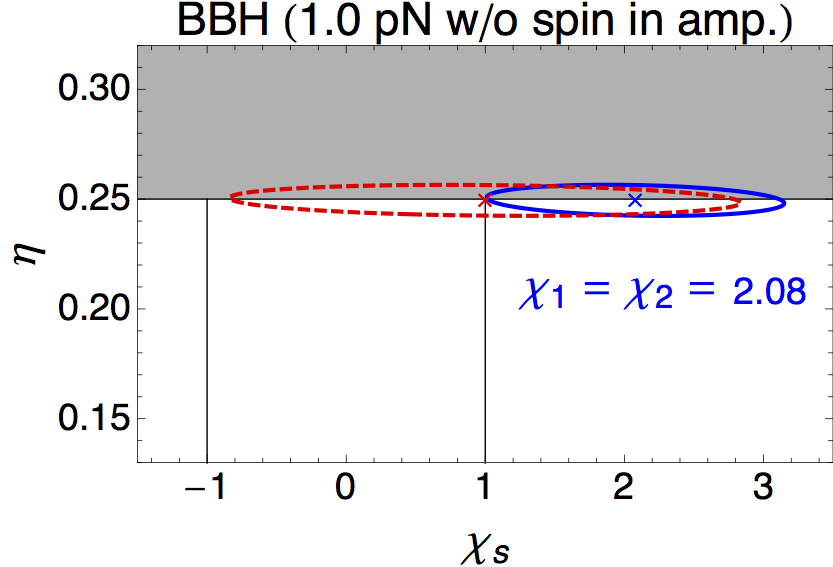}
	\end{minipage}
	
	\

	\begin{minipage}[b]{0.38\textwidth}
	\centering
		\includegraphics[width = \textwidth]{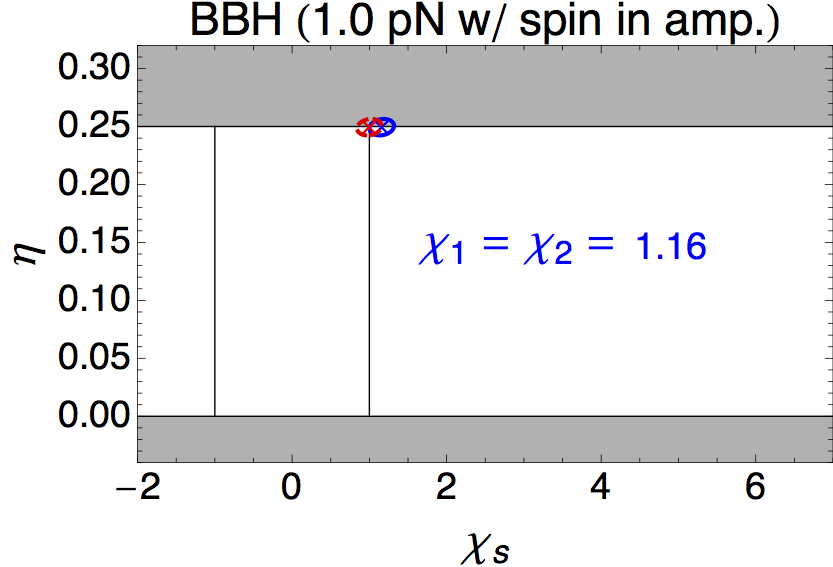}
	\end{minipage}
	~
	~
	\begin{minipage}[b]{0.38\textwidth}
	\centering
		\includegraphics[width = \textwidth]{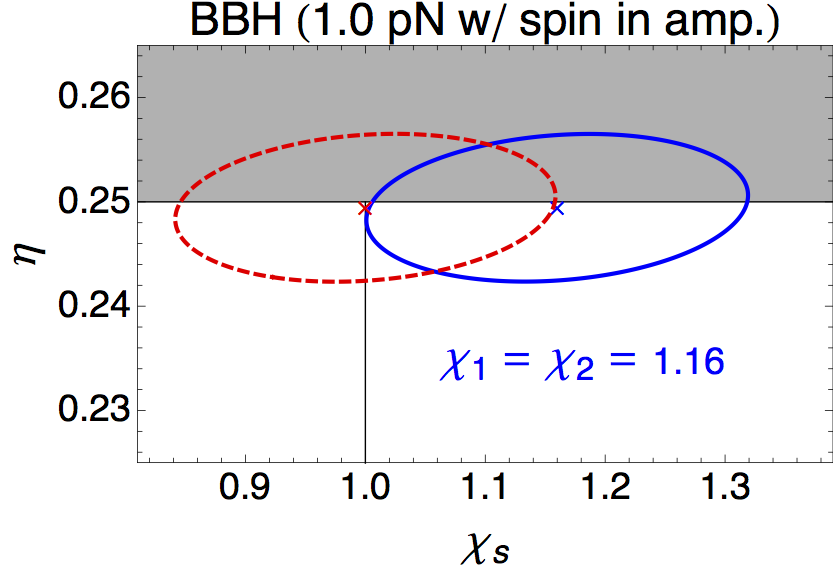}
	\end{minipage}
	\caption{\footnotesize These plots show 1$\sigma$ error ellipses in the $\eta-\chi_s$ parameter space for various spinning, amplitude-corrected waveforms with spin corrections in the phase to 2.5 pN order as described in Sec.~\ref{sec:waveforms}.  The title of each plot indicates the pN order amplitude correction.  For the 1.0 pN order amplitude-corrected waveform, the title also indicates whether spin corrections have been included in the amplitude.  These ellipses are calculated for a spinning BBH system with true parameters $m_1 = 10 \mbox{ M}_\odot$, $m_2 = 11 \mbox{ M}_\odot$, $t_c = 0$, $\phi_c = 0$, $\theta = \pi/6$, $\phi = \pi/6$, $\psi = \pi/4$ and $\iota = \pi/3$ and with a fixed SNR of $\rho = 10$.  The component spins for the solid, blue ellipses are given as an inlay on the plot.  These spins indicate the minimum detectable apparent violation of cosmic censorship.  The dashed, red ellipses are calculated with the fiducial spin values of $\chi_1 = \chi_2 = 1$ in each plot. The plots on the left are all to the same scale for comparison purposes.  The plots on the right are shown to a scale appropriate for each ellipse. The unphysical areas of parameter space are shaded gray.  The vertical solid lines bound the region of parameter space that is consistent with cosmic censorship ($-1 \le \chi_s \le 1$).}
	\label{fig:bbhspinplots}
\end{figure*}

\begin{figure*}[p]
	\begin{minipage}[b]{0.38\textwidth}
	\centering
		\includegraphics[width = \textwidth]{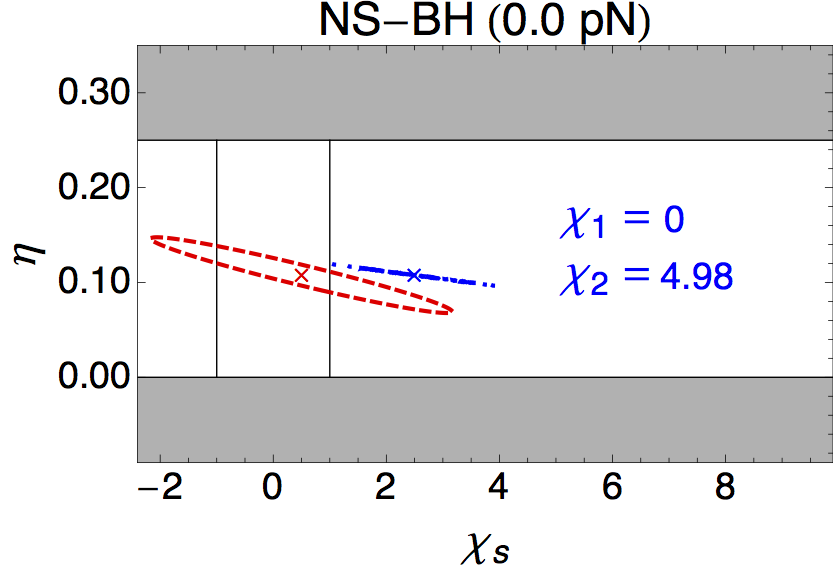}
	\end{minipage}
	~
	~
	\begin{minipage}[b]{0.38\textwidth}
	\centering
		\includegraphics[width = \textwidth]{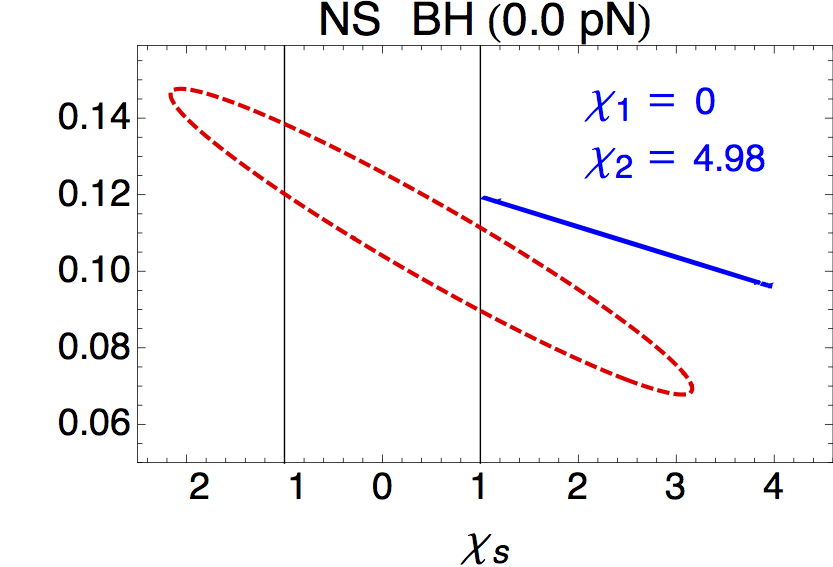}
	\end{minipage}
	
	\
	
	\begin{minipage}[b]{0.38\textwidth}
	\centering
		\includegraphics[width = \textwidth]{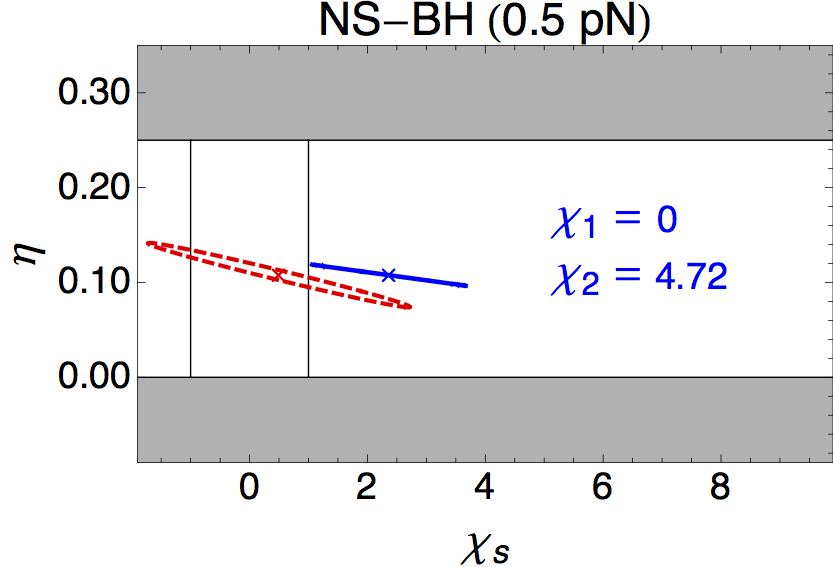}
	\end{minipage}
	~
	~
	\begin{minipage}[b]{0.38\textwidth}
	\centering
		\includegraphics[width = \textwidth]{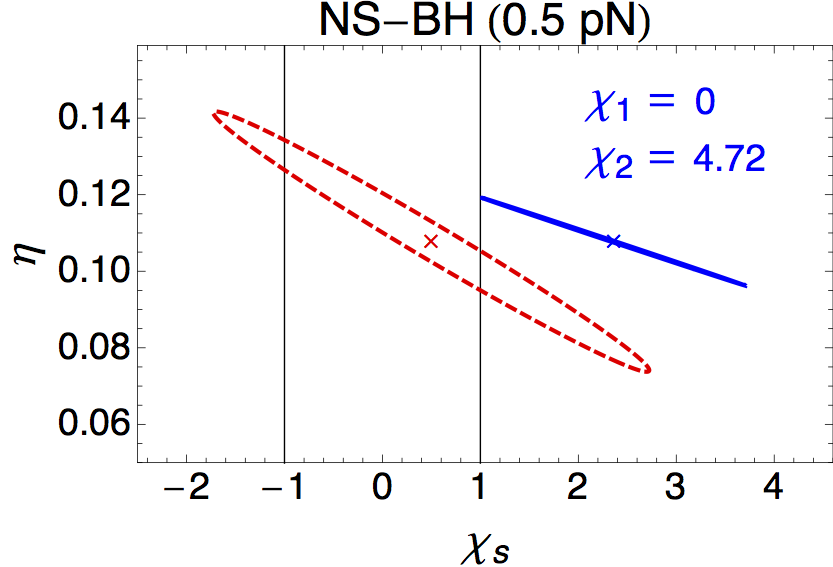}
	\end{minipage}
	
	\

	\begin{minipage}[b]{0.38\textwidth}
	\centering
		\includegraphics[width = \textwidth]{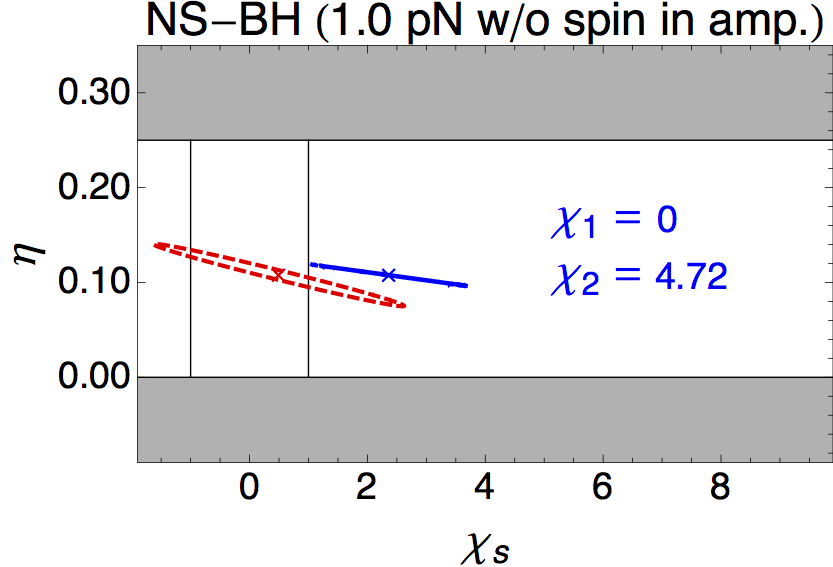}
	\end{minipage}
	~
	~
	\begin{minipage}[b]{0.38\textwidth}
	\centering
		\includegraphics[width = \textwidth]{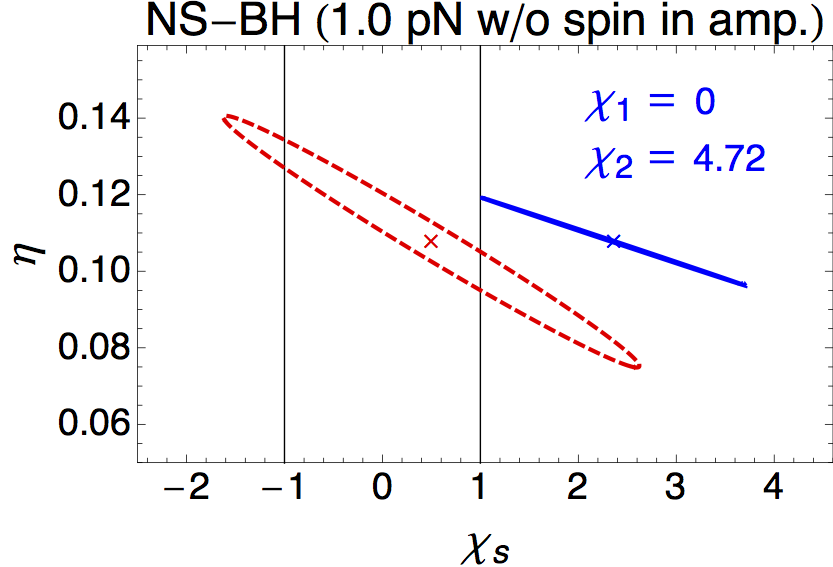}
	\end{minipage}
	
	\

	\begin{minipage}[b]{0.38\textwidth}
	\centering
		\includegraphics[width = \textwidth]{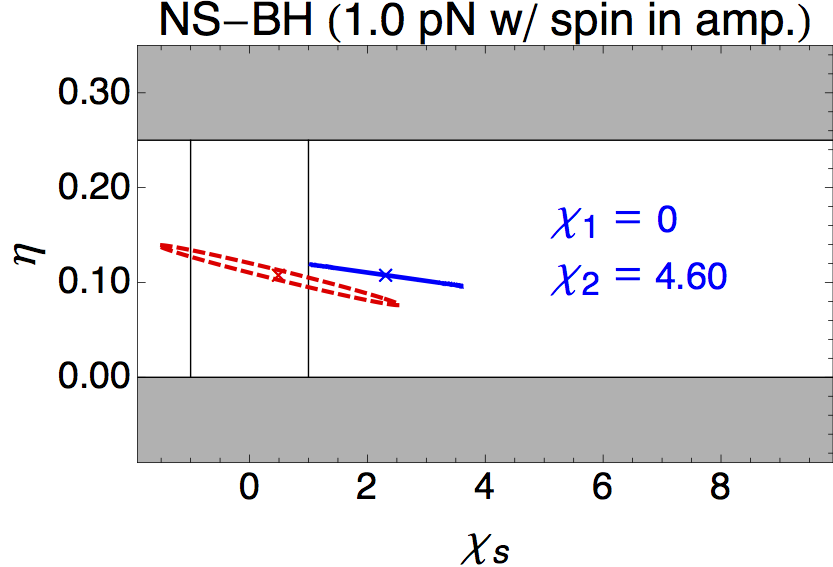}
	\end{minipage}
	~
	~
	\begin{minipage}[b]{0.38\textwidth}
	\centering
		\includegraphics[width = \textwidth]{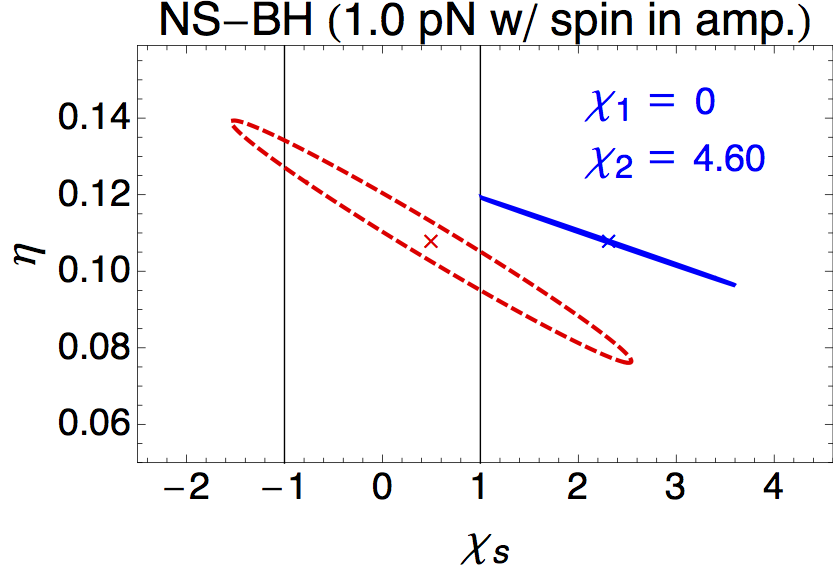}
	\end{minipage}
	\caption{\footnotesize These plots show 1$\sigma$ error ellipses in the $\eta-\chi_s$ parameter space for various spinning, amplitude-corrected waveforms with spin corrections in the phase to 2.5 pN order as described in Sec.~\ref{sec:waveforms}.  The title of each plot indicates the pN order amplitude correction.  For the 1.0 pN order amplitude-corrected waveform, the title also indicates whether spin corrections have been included in the amplitude.  These ellipses are calculated for a spinning NS-BH system with true parameters $m_1 = 1.4 \mbox{ M}_\odot$, $m_2 = 10 \mbox{ M}_\odot$, $t_c = 0$, $\phi_c = 0$, $\theta = \pi/6$, $\phi = \pi/6$, $\psi = \pi/4$ and $\iota = \pi/3$ and with a fixed SNR of $\rho = 10$.  The component spins for the solid, blue ellipses are given as an inlay on the plot.  These spins indicate the minimum detectable apparent violation of cosmic censorship.  The dashed, red ellipses are calculated with the fiducial spin values of $\chi_1 = 0$ and $\chi_2 = 1$ in each plot. The plots on the left are all to the same scale for comparison purposes.  The plots on the right are shown to a scale appropriate for each ellipse. The unphysical areas of parameter space are shaded gray.  The vertical solid lines bound the region of parameter space that is consistent with cosmic censorship ($-1 \le \chi_s \le 1$).}
	\label{fig:nsbhspinplots}
\end{figure*}

\begin{table*}[!]
	\caption{{\bf Spinning BBH System} {\footnotesize This table shows results for the spinning BBH system with true parameters $m_1 = 10 \mbox{ M}_\odot$, $m_2 = 11 \mbox{ M}_\odot$, $t_c = 0$, $\phi_c = 0$, $\theta = \pi/6$, $\phi = \pi/6$, $\psi = \pi/4$ and $\iota = \pi/3$ and with a fixed SNR of $\rho = 10$.  The spins $\chi_1$ and $\chi_2$ for each scenario are given in the tables.  The tables show the 1$\sigma$ measurement errors and correlation coefficients obtained from the Fisher matrix using spinning waveforms as described in Sec.~\ref{sec:waveforms} with spin corrections in the phase to 2.5 pN order.  Also given in the tables is the distance $D_{\rm M}$ of the system in order to achieve the fixed SNR of 10.  Different order amplitude corrections, with and without spin in the amplitude, are given in different rows of the tables.  The top table shows results for fiducial spin values of $\chi_1 = \chi_2 = 1$, and the bottom table shows results for the minimum detectable violating spins for each waveform.  The first row of the bottom table is for the minimum violating spin when the entire $\eta$ parameter space is considered, and the second row is when only the physical $\eta$ parameter space is considered.}}
	\bigskip
	\begin{minipage}[b]{1.0\textwidth}
	\centering
		\begin{tabular}{l l l r@{.} lr@{.} lr@{.} lr@{.} lr@{.} lr@{.}l }
		\hline
		pN order in amplitude \hspace{0.2pt}
		& $\chi_1 = \chi_2$ \hspace{0.2pt}
		& $D_{\rm M}$ (Mpc) \hspace{0.2pt}
		& \multicolumn{2}{c}{$\Delta \mathcal{M}/\mathcal{M}$} \hspace{0.2pt}
		& \multicolumn{2}{c}{$\Delta \eta$} \hspace{0.2pt}
		& \multicolumn{2}{c}{$\Delta \chi_a$} \hspace{0.2pt}
		& \multicolumn{2}{c}{$\Delta \chi_s$} \hspace{0.2pt}
		& \multicolumn{2}{c}{$c_{\eta\chi_s}$} \hspace{0.2pt}
		& \multicolumn{2}{c}{$c_{\mathcal{M}\chi_s}$} \\
		\hline
		\hline
		 0.0 pN  \hspace{0.2pt} & 1 \hspace{0.2pt} &  938 \hspace{0.2pt} & 0&0367 \hspace{0.2pt} & 2&06 \hspace{0.2pt} & 522&2 \hspace{0.2pt} & 30&8 \hspace{0.2pt} & --0&9998 \hspace{0.2pt} &  0&9989 \\
		 0.5 pN & 1 & 938  & 0&00420 & 0&0411& 42&2 & 2&66 & --0&3543 \hspace{0.2pt} & 0&9197 \\
		1.0 pN & 1 &  879 & 0&00328 & 0&00704 & 28&8 & 1&84 & --0&1541 \hspace{0.2pt} & 0&8777  \\
		1.5 pN & 1 & 879 & 0&00339  & 0&00807  & 30&5 & 1&94 & --0&1422 \hspace{0.2pt} & 0&8844 \\
		 2.0 pN & 1 & 851 & 0&00360 & 0&00752 & 34&3 & 2&18 & --0&1646 \hspace{0.2pt} & 0&8979  \\
		 2.5 pN & 1 & 851 & 0&00329 & 0&00766 & 29&4 & 1&87 & --0&1491 \hspace{0.2pt} & 0&8790  \\
		 \hline
		 1.0 pN + spin & 1 & 879  & 0&00164 & 0&00709  & 2&84 & 0&159 & 0&1568 & 0&3184 \\
		 1.5 pN + spin & 1 & 935 & 0&00168 & 0&00882 & 2&56 & 0&167 & 0&2238 & 0&3168 \\
		 2.0 pN + spin  & 1 & 901 & 0&00167 & 0&00809 & 2&44 & 0&159 & 0&1734 & 0&3169  \\
		 2.5 pN + spin & 1 & 902 & 0&00166 & 0&00825 & 2&44 & 0&159 & 0&1741 & 0&3201  \\
		 \hline
		\end{tabular}
	\end{minipage}

	\vspace{0.01\textwidth}

	\begin{minipage}[b]{1.0\textwidth}
	\centering
		\begin{tabular}{l r@{.}l l r@{.} lr@{.} lr@{.} lr@{.} lr@{.} lr@{.}l}
		\hline
		pN order in amplitude \hspace{0.2pt}
		  &   \multicolumn{2}{c}{$\chi_1=\chi_2$} \hspace{0.2pt}
		  &  $D_{\rm M}$ (Mpc)  \hspace{0.2pt}
		  &  \multicolumn{2}{c}{$\Delta \mathcal{M}/\mathcal{M}$}  \hspace{0.2pt}
		  &  \multicolumn{2}{c}{$\Delta \eta$} \hspace{0.2pt}
		  &  \multicolumn{2}{c}{$\Delta \chi_a$} \hspace{0.2pt}
		  &  \multicolumn{2}{c}{$\Delta \chi_s$} \hspace{0.2pt}
		  &  \multicolumn{2}{c}{$c_{\eta\chi_s}$} \hspace{0.2pt}
		  &  \multicolumn{2}{c}{$c_{\mathcal{M}\chi_s}$} \\
		 \hline
		 \hline
		 0.0 pN  (w/o $\eta$ bound)  \hspace{0.2pt} &  4&81 \hspace{0.2pt}  &  938 \hspace{0.2pt} &  0&00609 \hspace{0.2pt} &  0&129 \hspace{0.2pt} &  103&2 \hspace{0.2pt} &  3&81 \hspace{0.2pt} &  --0&9998 \hspace{0.2pt} &  --0&9452 \\
		 0.0 pN  (w/ $\eta$ bound)  & 1&43 & 938 & 0&00918 & 0&936 & 226&3 & 11&8 & --0&9994 & 0&9757  \\
		 0.5 pN & 2&33 & 938  & 0&00377 & 0&0411 & 23&6 & 1&51 & --0&5014 & 0&8082  \\
		 1.0 pN & 2&08 &  879 & 0&00304 & 0&00709 & 15&0 & 1&07 & --0&1851 & 0&8390  \\
		 1.5 pN & 2&12 & 879 & 0&00316  & 0&00810  & 15&7& 1&12 & --0&1737 & 0&8479  \\
		 2.0 pN & 2&21 & 850 & 0&00335 & 0&00755 & 16&7 & 1&21 & --0&1995 & 0&8652 \\
		 2.5 pN & 2&09 & 851 & 0&00307 & 0&00769 & 15&2 & 1&09 & --0&1809 & 0&8409 \\
		 \hline
		 1.0 pN + spin  & 1&16 & 879  & 0&00165 & 0&00708  & 2&40 & 0&159 & 0&1693 & 0&3344  \\
		 1.5 pN + spin  & 1&18 & 945 & 0&00170 & 0&00891 & 2&57 & 0&172 & 0&2471 & 0&3383 \\	
		 2.0 pN + spin & 1&17 & 909 & 0&00169 & 0&00815 & 2&45 & 0&163 & 0&1891 & 0&3339 \\
		 2.5 pN + spin & 1&17 & 909 & 0&00168 & 0&00831 & 2&45 & 0&164 & 0&1916 & 0&3381 \\	
		\hline
		\end{tabular}
	\end{minipage}
	\label{tab:bbhspintable}

	\caption{{\bf Spinning BBH System with 3.0 pN and 3.5 pN Spin-Orbit Phase Terms} \footnotesize This table shows results for the spinning BBH system with true parameters $m_1 = 10 \mbox{ M}_\odot$, $m_2 = 11 \mbox{ M}_\odot$, $t_c = 0$, $\phi_c = 0$, $\chi_1 = \chi_2 = 1$, $\theta = \pi/6$, $\phi = \pi/6$, $\psi = \pi/4$ and $\iota = \pi/3$ and with a fixed SNR of $\rho = 10$.  The table shows the 1$\sigma$ measurement errors and correlation coefficients obtained from the Fisher matrix using spinning waveforms as described in Sec.~\ref{sec:waveforms} with spin corrections in the phase to 3.5 pN order.  Also given in the table is the distance $D_{\rm M}$ of the system in order to achieve the fixed SNR of 10.   Different order amplitude corrections, with and without spin in the amplitude, are given in different rows of the table.}
\label{tab:newspintermstableBBH}
\bigskip
\centering
		\begin{tabular}{l l l r@{.} lr@{.} lr@{.} lr@{.} lr@{.} lr@{.}l }
		\hline
		pN order in amplitude \hspace{0.2pt}
		& $\chi_1 = \chi_2$ \hspace{0.2pt}
		& $D_{\rm M}$ (Mpc) \hspace{0.2pt}
		& \multicolumn{2}{c}{$\Delta \mathcal{M}/\mathcal{M}$} \hspace{0.2pt}
		& \multicolumn{2}{c}{$\Delta \eta$} \hspace{0.2pt}
		& \multicolumn{2}{c}{$\Delta \chi_a$} \hspace{0.2pt}
		& \multicolumn{2}{c}{$\Delta \chi_s$} \hspace{0.2pt}
		& \multicolumn{2}{c}{$c_{\eta\chi_s}$} \hspace{0.2pt}
		& \multicolumn{2}{c}{$c_{\mathcal{M}\chi_s}$} \\
		\hline
		\hline
		 0.0 pN  \hspace{0.2pt} &  1  &  938 \hspace{0.2pt} &  0&00431 \hspace{0.2pt} &  0&0681 \hspace{0.2pt} &  14&9 \hspace{0.2pt} &  1&09 \hspace{0.2pt} &  0&9173 \hspace{0.2pt} &  0&8811 \\
		 0.5 pN & 1  & 938  & 0&00419 & 0&0358 & 20&9 & 1&14 & 0&1460 & 0&9062  \\
		 1.0 pN & 1  &  879 & 0&00309 & 0&00697 & 16&8 & 1&13 & 0&1315 & 0&8686  \\
		 1.5 pN  & 1 & 879 & 0&00324  & 0&00798  & 17&5 & 1&17 & 0&1155 & 0&8788  \\
		 2.0 pN & 1 & 850 & 0&00356 & 0&00742 & 19&4 & 1&30 & 0&1271 & 0&9000 \\
		 2.5 pN & 1 & 851 & 0&00314 & 0&00757 & 17&4 & 1&16 & 0&1237 & 0&8736 \\
		 \hline
		 1.0 pN + spin   & 1  & 879  & 0&00165 & 0&00707  & 2&42 & 0&166 & 0&1995 & 0&4112  \\
		 1.5 pN + spin & 1 & 935 & 0&00172 & 0&00880 & 2&55 & 0&176 & 0&2628 & 0&4319 \\
		 2.0 pN + spin & 1 & 901 & 0&00172 & 0&00806 & 2&44 & 0&168 & 0&2106 & 0&4303 \\
		 2.5 pN + spin & 1 & 902 & 0&00169 & 0&00822 & 2&43 & 0&168 & 0&2140 & 0&4224 \\
		 \hline
		\end{tabular}
\end{table*}

\begin{table*}[!]
	\caption{{\bf Spinning NS-BH System} \footnotesize This table shows results for the spinning NS-BH system with true parameters $m_1 = 1.4 \mbox{ M}_\odot$, $m_2 = 10 \mbox{ M}_\odot$, $t_c = 0$, $\phi_c = 0$, $\theta = \pi/6$, $\phi = \pi/6$, $\psi = \pi/4$ and $\iota = \pi/3$ and with a fixed SNR of $\rho = 10$.  The spins $\chi_1$ and $\chi_2$ for each scenario are given in the tables.  The tables show the 1$\sigma$ measurement errors and correlation coefficients obtained from the Fisher matrix using spinning waveforms as described in Sec.~\ref{sec:waveforms} with spin corrections in the phase to 2.5 pN order.  Also given in the table is the distance $D_{\rm M}$ of the system in order to achieve the fixed SNR of 10.   Different order amplitude corrections, with and without spin in the amplitude, are given in different rows of the tables.  The top table shows results for fiducial spin values of $\chi_1 = 0$ and $\chi_2 = 1$, and the bottom table shows results for the minimum detectable violating black hole spin for each waveform.  The first row of the bottom table is for the minimum violating spin when the entire $\eta$ parameter space is considered, and the second row of the bottom table is for the minimum violating spin when only the physical $\eta$ parameter space is considered.}
	\bigskip
	\begin{minipage}[b]{1.0\textwidth}
	\centering
		\begin{tabular}{l l l l r@{.}l r@{.}l r@{.}l r@{.}l r@{.}l r@{.}l}
		\hline
		  pN order in amplitude \hspace{0.2pt}
		  &  $\chi_1$ \hspace{0.2pt}
		  &   $\chi_2$  \hspace{0.2pt}
		  &  $D_{\rm M}$ (Mpc)  \hspace{0.2pt}
		  &  \multicolumn{2}{c}{$\Delta \mathcal{M}/\mathcal{M}$} \hspace{0.2pt}
		  &  \multicolumn{2}{c}{$\Delta \eta$} \hspace{0.2pt}
		  &  \multicolumn{2}{c}{$\Delta \chi_a$} \hspace{0.2pt}
		  &  \multicolumn{2}{c}{$\Delta \chi_s$} \hspace{0.2pt}
		  &  \multicolumn{2}{c}{$c_{\eta\chi_s}$} \hspace{0.2pt}
		  &  \multicolumn{2}{c}{$c_{\mathcal{M}\chi_s}$} \\
		\hline
		\hline
		 0.0 pN  \hspace{0.2pt} &  0 \hspace{0.2pt} &  1 \hspace{0.2pt}  &  383 \hspace{0.2pt} &  0&00199 \hspace{0.2pt} &  0&0399 \hspace{0.2pt} &  3&30 \hspace{0.2pt} &  2&66 \hspace{0.2pt} &  --0&9929 \hspace{0.2pt} &  0&9981 \\
		 0.5 pN & 0 & 1 & 391  & 0&00164 & 0&0340 & 2&76 & 2&22 & --0&9879 & 0&9964  \\
		 1.0 pN & 0 & 1 &  364 & 0&00156 & 0&0329 & 2&64 & 2&12 & --0&9875 & 0&9965  \\
		 1.5 pN  & 0 & 1 & 361 & 0&00159  & 0&0335  & 2&68 & 2&16 & --0&9869 & 0&9963  \\
		 2.0 pN & 0 & 1 & 356 & 0&00159 & 0&0336 & 2&68 & 2&16 & --0&9474 & 0&9965  \\
		 2.5 pN & 0 & 1 & 355 & 0&00158 & 0&0336 & 2&68 & 2&15 & --0&9868 & 0&9964  \\
		 \hline
		 1.0 pN + spin   & 0 & 1 & 363  & 0&00150 & 0&0316  & 2&53 & 2&04 & --0&9864 & 0&9962  \\
		 1.5 pN + spin & 0 & 1 & 376 & 0&00154 & 0&0322 & 2&59 & 2&09 & --0&9859 & 0&9958  \\
		 2.0 pN + spin & 0 & 1 & 371 & 0&00154 & 0&0323 & 2&59 & 2&08 & --0&9861 & 0&9960  \\
		 2.5 pN + spin & 0 & 1 & 370 & 0&00154 & 0&0323 & 2&59 & 2&08 & --0&9859 & 0&9959  \\
		 \hline
		\end{tabular}
	\end{minipage}

	\vspace{0.01\textwidth}

	\begin{minipage}[b]{1.0\textwidth}
	\centering
		\begin{tabular}{l l r@{.}l l r@{.}l r@{.}l r@{.}l r@{.}l r@{.}l r@{.}l}
		\hline
		  pN order in amplitude \hspace{0.2pt}
		  &  $\chi_1$  \hspace{0.2pt}
		  &   \multicolumn{2}{c}{$\chi_2$} \hspace{0.2pt}
		  &  $D_{\rm M}$ (Mpc)  \hspace{0.2pt}
		  &  \multicolumn{2}{c}{$\Delta \mathcal{M}/\mathcal{M}$} \hspace{0.2pt}
		  &  \multicolumn{2}{c}{$\Delta \eta$} \hspace{0.2pt}
		  &  \multicolumn{2}{c}{$\Delta \chi_a$} \hspace{0.2pt}
		  &  \multicolumn{2}{c}{$\Delta \chi_s$} \hspace{0.2pt}
		  &  \multicolumn{2}{c}{$c_{\eta\chi_s}$} \hspace{0.2pt}
		  &  \multicolumn{2}{c}{$c_{\mathcal{M}\chi_s}$} \\
		\hline
		\hline
		 0.0  pN \hspace{0.2pt} &  0 \hspace{0.2pt} &  4&98  \hspace{0.2pt} &  383 \hspace{0.2pt} &  0&000469 \hspace{0.2pt} &  0&0118 \hspace{0.2pt} &  2&33 \hspace{0.2pt} &  1&49 \hspace{0.2pt} &  --0&99996   \hspace{0.2pt} &  --0&7490  \\
		 0.5 pN & 0 & 4&72 & 391  & 0&000381 & 0&0115 & 2&11 & 1&35 & --0&9998 & --0&6461  \\
		 1.0 pN & 0 & 4&72 &  364 & 0&000357 & 0&0115 & 2&11 & 1&35 & --0&9998 & --0&6148  \\
		 1.5  pN & 0 & 4&78 & 362 & 0&000359  & 0&0115  & 2&14 & 1&37 & --0&9998 & --0&6162  \\
		 2.0  pN & 0 & 4&84 & 358 & 0&000361 & 0&0116 & 2&19 & 1&41 & --0&9998 & --0&6277  \\
		 2.5 pN & 0 & 4&84 & 356 & 0&000360 & 0&0117 & 2&21 & 1&42 & --0&9998 & --0&6232  \\
		 \hline
		 1.0 pN + spin   & 0 & 4&60 & 365  & 0&000337 & 0&0115  & 2&03 & 1&30 & --0&9997 & --0&5566  \\
		 1.5 pN + spin & 0 & 4&52 & 435 & 0&000382 & 0&0114 & 1&97 & 1&26 & --0&9995 & --0&6175  \\
		 2.0 pN + spin & 0 & 4&50 & 430 & 0&000372 & 0&0113 & 1&94 & 1&24 & --0&9995 & --0&5960  \\
		 2.5 pN + spin & 0 & 4&50 & 429 & 0&000371 & 0&0113 & 1&94 & 1&24 & --0&9994 & --0&5914  \\
		\hline
		\end{tabular}
	\end{minipage}
\label{tab:nsbhspintable}

	\caption{{\bf Spinning NS-BH System with 3.0 pN and 3.5 pN Spin-Orbit Phase Terms} \footnotesize This table shows results for the spinning NS-BH system with true parameters $m_1 = 1.4 \mbox{ M}_\odot$, $m_2 = 10 \mbox{ M}_\odot$, $t_c = 0$, $\phi_c = 0$, $\chi_1 = 0$, $\chi_2 = 1$, $\theta = \pi/6$, $\phi = \pi/6$, $\psi = \pi/4$ and $\iota = \pi/3$ and with a fixed SNR of $\rho = 10$.  The table shows the 1$\sigma$ measurement errors and correlation coefficients obtained from the Fisher matrix using spinning waveforms as described in Sec.~\ref{sec:waveforms} with spin corrections in the phase to 3.5 pN order.  Also given in the table is the distance $D_{\rm M}$ of the system in order to achieve the fixed SNR of 10.   Different order amplitude corrections, with and without spin in the amplitude, are given in different rows of the table.}
\label{tab:newspintermstableNSBH}
\bigskip
\centering
		\begin{tabular}{l l l l r@{.}l r@{.}l r@{.}l r@{.}l r@{.}l r@{.}l}
		\hline
		  pN order in amplitude \hspace{0.2pt}
		  &  $\chi_1$ \hspace{0.2pt}
		  &   $\chi_2$  \hspace{0.2pt}
		  &  $D_{\rm M}$ (Mpc)  \hspace{0.2pt}
		  &  \multicolumn{2}{c}{$\Delta \mathcal{M}/\mathcal{M}$} \hspace{0.2pt}
		  &  \multicolumn{2}{c}{$\Delta \eta$} \hspace{0.2pt}
		  &  \multicolumn{2}{c}{$\Delta \chi_a$} \hspace{0.2pt}
		  &  \multicolumn{2}{c}{$\Delta \chi_s$} \hspace{0.2pt}
		  &  \multicolumn{2}{c}{$c_{\eta\chi_s}$} \hspace{0.2pt}
		  &  \multicolumn{2}{c}{$c_{\mathcal{M}\chi_s}$} \\
		\hline
		\hline
		 0.0  pN \hspace{0.2pt} &  0 \hspace{0.2pt} &  1  \hspace{0.2pt} &  383 \hspace{0.2pt} &  0&000452 \hspace{0.2pt} &  0&00527 \hspace{0.2pt} &  1&05 \hspace{0.2pt} &  0&900 \hspace{0.2pt} &  0&8153 \hspace{0.2pt} &  0&9208  \\
		 0.5 pN & 0 & 1 & 390  & 0&000396 & 0&00480 & 0&956 & 0&822 & 0&7442 & 0&9138  \\
		 1.0 pN & 0 & 1 &  364 & 0&000387 & 0&00462 & 0&968 & 0&831 & 0&7350 & 0&9181  \\
		 1.5 pN & 0 & 1 & 361 & 0&000394 & 0&00469 & 0&991 & 0&850 & 0&7251 & 0&9192 \\
		 2.0 pN & 0 & 1 & 356 & 0&000397 & 0&00465 & 1&01 & 0&868 & 0&7222 & 0&9222 \\
		 2.5 pN & 0 & 1 & 355 & 0&000399 & 0&00466 & 1&02 & 0&876 & 0&7182 & 0&9229 \\
		 \hline
		 1.0 pN + spin & 0 & 1 & 363 & 0&000385 & 0&00461  & 0&963 & 0&827 & 0&7324 & 0&9173  \\
		 1.5 pN + spin & 0 & 1 & 376 & 0&000390 & 0&00474 & 0&954 & 0&819 & 0&7295 & 0&9137 \\
		 2.0 pN + spin & 0 & 1 & 371 & 0&000392 & 0&00471 & 0&969 & 0&832 & 0&7293 & 0&9162 \\
		 2.5 pN + spin & 0 & 1 & 370 & 0&000393 & 0&00472 & 0&976 & 0&838 & 0&7251 & 0&9167 \\
		\hline
		\end{tabular}
\end{table*}

Fig.~\ref{fig:spinetabound} compares the minimally violating spin values for a Newtonian-amplitude waveform when considering the entire parameter space (left plot) versus considering only the physical area of parameter space (right plot). The error ellipse on the right of Fig.~\ref{fig:spinetabound} is consistent with the Kerr bound when considering the entire parameter space, but it is inconsistent with the Kerr bound when considering only the area of the ellipse within the physically-allowed region of $\eta$.  Results are shown for only the spinning BBH system.  The spinning NS-BH system is not affected by bounding values of $\eta$ due to the error ellipse's orientation and placement in parameter space, as is evident in Fig.~\ref{fig:nsbhspinplots}.  

The strong correlation between the symmetric mass ratio $\eta$ and spin when using a Newtonian-amplitude waveform has been studied by \cite{spinmassdegen} and \cite{Alex}, among others.  The correlation between mass and spin can be seen in Fig.s~\ref{fig:spinetabound},~\ref{fig:bbhspinplots} and~\ref{fig:nsbhspinplots}.  As a result, the spin parameters are not well measured with the Newtonian-amplitude waveform when considering the full $\eta-\chi_s$ parameter space.  However, by restricting the parameter space to only the physical region of $\eta$ for the spinning BBH system, aLIGO's ability to detect violations of the Kerr bound increases by about a factor of three.  This result is also summarized in Table~\ref{tab:bbhspintable}. 

We examine how the measurability of spin is affected by including spin-independent and spin-dependent amplitude corrections.  The measurability of spin for waveforms with spin corrections included in the phase but only spin-independent amplitude corrections was reported in \cite{Chris}.  Since then, more accurate spin corrections to the phase and spin corrections to the amplitude have been calculated. As described in Sec.~\ref{sec:waveforms}, here we use the waveforms given in \cite{Evan}, which include spin corrections in the amplitude to 2.0 pN order and spin corrections in the phase to 2.5 pN order.  Later in this Sec. we address the more recent spin-orbit corrections at 3.0 pN and 3.5 pN order in the phase.  

The results for different order amplitude corrections, with and without spin corrections in the amplitude, are shown in Fig.~\ref{fig:bbhspinplots} and Table~\ref{tab:bbhspintable} for the BBH system and in Fig.~\ref{fig:nsbhspinplots} and Table~\ref{tab:nsbhspintable} for the NS-BH system.  The plots in Fig.s~\ref{fig:bbhspinplots} and~\ref{fig:nsbhspinplots} show $1\sigma$ error ellipses for fiducial spin values of $\chi_1 = \chi_2 =1 $ for the BBH system and $\chi_1 = 0$, $\chi_2 = 1$ for the NS-BH system (red, dashed ellipses).  In addition, the plots show $1\sigma$ error ellipses for the minimum detectable violation of the Kerr bound (blue, solid ellipses).  The top table in Tables~\ref{tab:bbhspintable} and~\ref{tab:nsbhspintable} show parameter root-mean-square errors and correlation coefficients for the fiducial spin values.  The bottom table in Tables~\ref{tab:bbhspintable} and~\ref{tab:nsbhspintable} show parameter errors and correlation coefficients for the systems that provide the minimum detectable violation of the Kerr bound with each waveform.

The BBH system is strongly affected by including amplitude corrections in the waveform and spin corrections in the amplitude.  There is about a factor-of-ten improvement in the measurability of the spin parameters when the lowest-order amplitude correction (0.5 pN) is included in the waveform and the spin terms in the phase are kept to 2.5 pN order.  Van Den Broeck and Sengupta also report on improved measurability of spin when amplitude corrections are included in the waveform in \cite{Chris}.  A notable effect in our calculations is that the symmetric mass ratio decouples from spin and many other waveform parameters when the first-order amplitude correction is included in the waveform.\footnote{Since the correlation between symmetric mass ratio and spin is decreased when using amplitude-corrected waveforms, restricting the error ellipse to only the physical area of $\eta$ parameter space does not significantly improve aLIGO's ability to detect apparent violations of cosmic censorship with these waveforms.  This is evident from the plots in Fig.~\ref{fig:bbhspinplots}.}  There is additional improvement in spin and mass measurability when the 1.0 pN order amplitude correction is included in the waveform, and there is a slight decrease in the degeneracy between spin and chirp mass for this waveform.  Furthermore, when spin corrections are included in the amplitude, which occurs at lowest-order at 1.0 pN, the measurability of spin improves by an additional factor of about ten.  In this case, including spin corrections in the amplitude breaks the correlation between chirp mass and spin even further.

The spinning NS-BH system is not strongly affected by including spin-dependent or nonspinning amplitude corrections.  There is a slight improvement in parameter measurability when moving from the Newtonian-amplitude waveform to the amplitude-corrected waveform, but this effect is not nearly as significant as with the spinning BBH system.  There is an even less significant improvement in parameter measurability for the spinning NS-BH system when moving from nonspinning amplitude corrections to spin-dependent amplitude corrections.  Overall, the Newtonian-amplitude spinning NS-BH waveform performs equally well as the amplitude-corrected waveforms when it comes to parameter measurability.

We do a brief study of how parameter measurability is affected by the 3.0 pN and 3.5 pN order spin-orbit corrections to the phase \cite{Bohe}.  Tables~\ref{tab:newspintermstableBBH} and~\ref{tab:newspintermstableNSBH} show the 1$\sigma$ errors and correlation coefficients from the Fisher matrix for the spinning BBH system and the spinning NS-BH system, respectively, with spin corrections in the phase to 3.5 pN order and the amplitude corrections varied as described in the table.  

The BBH system and the NS-BH system are both affected in some way by the 3.0 pN and 3.5 pN order spin-orbit terms in the phase.  For the BBH system, there is more than a factor-of-ten improvement in the symmetric mass ratio and spin parameter measurability for the Newtonian-amplitude waveform, and there is about a factor-of-ten improvement in the chirp mass measurability.  The degeneracy between the chirp mass and the spin is slightly decreased when the 3.0 pN and 3.5 pN order spin terms in the phase are included in the Newtonian-amplitude waveform, which may be what leads to the improved measurability of mass and spin.  The amplitude-corrected waveforms without spin in the amplitude show improved measurability of about a factor of two for the spin but not the mass parameters, and the amplitude-corrected waveforms with spin corrections in the amplitude are minimally affected by the  3.0 pN and 3.5 pN order spin-orbit corrections to the phase.  

For all of the different amplitude-corrected waveforms, the spinning NS-BH system shows about a factor-of-ten improvement in the measurability of the mass parameters when the 3.0 pN and 3.5 pN order spin-orbit corrections are included in the phase and about a factor of three improvement in the measurability of the spin parameters.  The 3.0 pN and 3.5 pN order spin-orbit phase corrections decrease the degeneracy between the symmetric mass ratio and the spin parameters, which may lead to the improved parameter measurability in this case.

\subsection{Detectable deviations from the no-hair theorem}
\label{sec:tidalresults}

\begin{figure*}[t]
	\begin{minipage}[b]{0.38\textwidth}
	\centering
		\includegraphics[width = \textwidth]{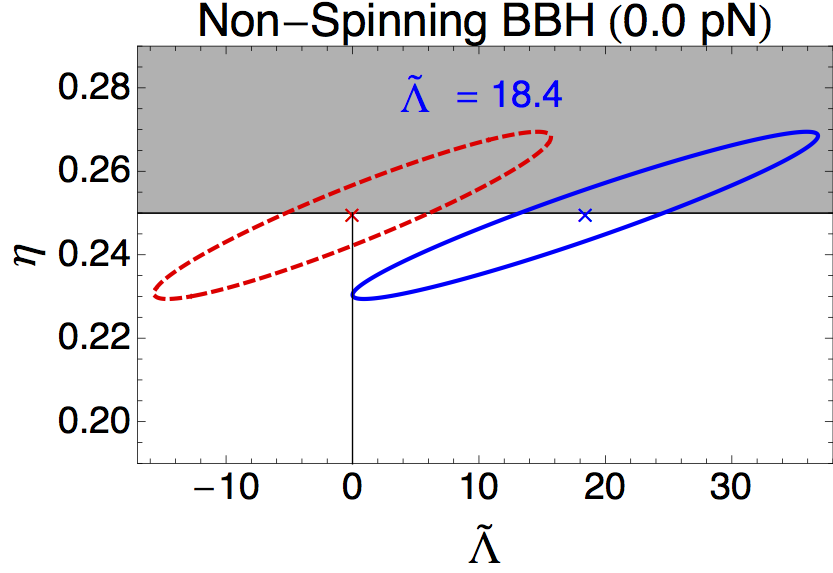}
	\end{minipage}	
	~
	~	
	\begin{minipage}[b]{0.38\textwidth}
	\centering
		\includegraphics[width = \textwidth]{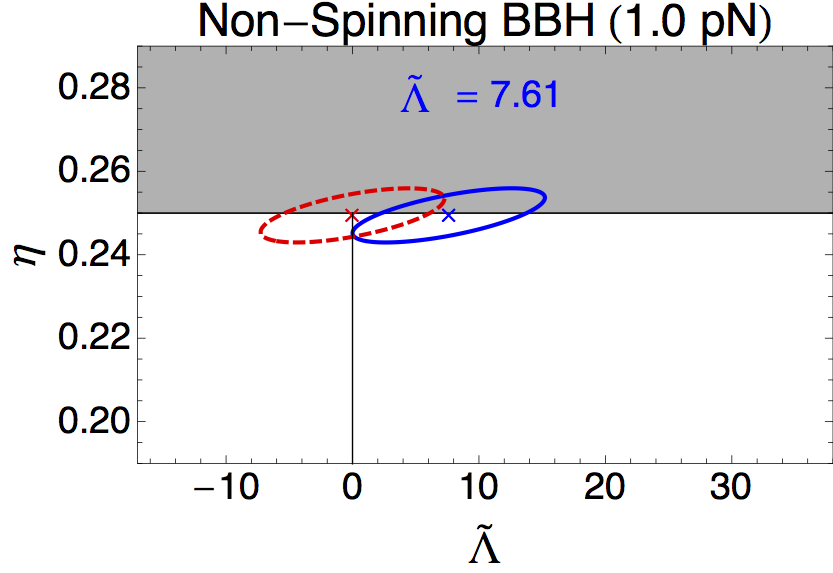}
	\end{minipage}
	\caption{\footnotesize These plots show 1$\sigma$ error ellipses in the $\eta-\tilde\Lambda$ parameter space for nonspinning, tidal waveforms, as described in Sec.~\ref{sec:waveforms}.  The title of each plot indicates the pN order amplitude correction.  These ellipses were calculated for a BBH system with true parameters $m_1 = 10 \mbox{ M}_\odot$, $m_2 = 11 \mbox{ M}_\odot$, $t_c = 0$, $\phi_c = 0$, $\theta = \pi/6$, $\phi = \pi/6$, $\psi = \pi/4$ and $\iota = \pi/3$ and with a fixed SNR of $\rho = 10$.  The tidal parameter $\tilde \Lambda$ for the solid, blue ellipses is given as an inlay on the plot.  These tidal deformability parameters indicate the minimum detectable deviation from the no-hair theorem for each waveform.  The dashed, red ellipses are shown for the fiducial tidal value of $ \tilde \Lambda = 0$ in each plot. The unphysical areas of parameter space are shaded gray.  The $\tilde \Lambda = 0$ axis indicates the area of parameter space that is consistent with the no-hair theorem.  Both plots above are shown to the same scale for comparison purposes.}
	\label{fig:bbhtidalplots}
\end{figure*}

\begin{table*}[!]
	\caption{{\bf Nonspinning Tidal BBH System} \footnotesize This table shows results for the nonspinning BBH system with true parameters $m_1 = 10 \mbox{ M}_\odot$, $m_2 = 11 \mbox{ M}_\odot$, $t_c = 0$, $\phi_c = 0$, $\theta = \pi/6$, $\phi = \pi/6$, $\psi = \pi/4$ and $\iota = \pi/3$ and with a fixed SNR of $\rho = 10$.  The tidal deformability parameter $\tilde \Lambda$ for each scenario is given in the tables.  The tables show the 1$\sigma$ measurement errors and correlation coefficients obtained from the Fisher matrix using the tidal waveform as described in Sec.~\ref{sec:waveforms}.  Also given in the table is the distance $D_{\rm M}$ of the system in order to achieve the fixed SNR of 10.  The phase is kept out to 3.5 pN order and different order amplitude corrections are given in different rows of the tables.  The top table shows results for a fiducial tidal parameter value of $\tilde \Lambda = 0$, and the bottom table shows results for the minimum detectable violating tides for each waveform.}
	\bigskip
	\begin{minipage}[b]{1.0\textwidth}
	\centering
		\begin{tabular}{l c c r@{.}l r@{.}l r@{.}l r@{.}l r@{.}l}
		\hline
		pN order in amplitude \hspace{0.2pt}
		&  $\tilde \Lambda$ \hspace{0.2pt}
		&  $D_{\rm M}$ (Mpc) \hspace{0.2pt}
		&  \multicolumn{2}{c}{$\Delta \mathcal{M}/\mathcal{M}$} \hspace{0.2pt}
		&   \multicolumn{2}{c}{$\Delta \eta$} \hspace{0.2pt}
		&   \multicolumn{2}{c}{$\Delta \tilde \Lambda$} \hspace{0.2pt}
		&  \multicolumn{2}{c}{$c_{\eta\tilde\Lambda}$} \hspace{0.2pt}
		&  \multicolumn{2}{c}{$c_{\mathcal{M}\tilde\Lambda}$}  \\
		\hline
		\hline
		 0.0 pN \hspace{0.2pt}  &  0 \hspace{0.2pt} &  938 \hspace{0.2pt} &  0&00281 \hspace{0.2pt} &  0&0200 \hspace{0.2pt} & 15&7 \hspace{0.2pt} &  0&9326  \hspace{0.2pt} &  0&8305 \\
		 0.5 pN & 0 \hspace{0.2pt} & 938  & 0&00234 & 0&0102 & 13&1 & 0&9019 & 0&7575  \\
		 1.0 pN & 0 \hspace{0.2pt} &  879 & 0&00115 & 0&00649 & 7&26 & 0&6090 & 0&1241  \\
		 1.5 pN & 0 \hspace{0.2pt} &  879 & 0&00123 & 0&00723 & 7&64 & 0&6569 & 0&2192  \\
		 2.0 pN & 0 \hspace{0.2pt} &  851 & 0&00118 & 0&00694 & 7&77 & 0&6375 & 0&1778  \\
		 2.5 pN & 0 \hspace{0.2pt} &  851 & 0&00118 & 0&00693 & 7&71 & 0&6331 & 0&1758  \\
		 \hline
		\end{tabular}
	\end{minipage}

	\vspace{0.01\textwidth}

	\begin{minipage}[b]{1.0\textwidth}
		\begin{tabular}{l r@{.}l c r@{.}l r@{.}l r@{.}l r@{.}l r@{.}l}
		\hline
		pN order in amplitude \hspace{0.2pt}
		&  \multicolumn{2}{c}{$\tilde \Lambda$} \hspace{0.2pt}
		&  $D_{\rm M}$ (Mpc) \hspace{0.2pt}
		&  \multicolumn{2}{c}{$\Delta \mathcal{M}/\mathcal{M}$} \hspace{0.2pt}
		&   \multicolumn{2}{c}{$\Delta \eta$} \hspace{0.2pt}
		&   \multicolumn{2}{c}{$\Delta \tilde \Lambda$} \hspace{0.2pt}
		&  \multicolumn{2}{c}{$c_{\eta\tilde\Lambda}$} \hspace{0.2pt}
		&  \multicolumn{2}{c}{$c_{\mathcal{M}\tilde\Lambda}$}  \\
		\hline
		\hline
		 0.0 pN  \hspace{0.2pt} &  18&4 \hspace{0.2pt} &  938 \hspace{0.2pt} &  0&00281 \hspace{0.2pt} &  0&0200 \hspace{0.2pt} &  18&4 \hspace{0.2pt} &  0&9512  \hspace{0.2pt} &  0&8581  \\
		 0.5 pN & 14&4 & 938  & 0&00232 & 0&0160 & 14&3 & 0&9193 & 0&7809  \\
		 1.0 pN & 7&61 &  879 & 0&00115 & 0&00649 & 7&61 & 0&6537 & 0&1723  \\
		 1.5 pN & 8&13 &  879 & 0&00123 & 0&00730 & 8&12 & 0&7039 & 0&2747  \\
		 2.0 pN & 8&19 & 850 & 0&00118 & 0&00698 & 8&18 & 0&6831 & 0&2308  \\
		 2.5 pN & 8&16 &  851 & 0&00118 & 0&00697 & 8&15 & 0&6811 & 0&2305  \\
		\hline
		\end{tabular}
	\end{minipage}	
\label{tab:bbhtidaltable}
\end{table*}

In this Sec. we discuss aLIGO's ability to detect deviations from the no-hair theorem using nonspinning, tidal waveforms, as described in Sec.~\ref{sec:waveforms}.  We keep the phase to 5.0 pN order, where point particle calculations are known to 3.5 pN order and the leading order tidal correction appears at 5.0 pN order.  We vary the amplitude corrections from 0.0 pN to 2.5 pN order.  We do not include any tidal corrections in the amplitude of the waveform, since they are not yet calculated.  We only investigate heavy systems, nominally BBH systems, without spin.  We look at a near equal mass BBH system with component masses $m_1 = 10$ ${\rm M}_\odot$ and $m_2 = 11$ ${\rm M}_\odot$.  As with the spinning system, the exactly equal mass limit is avoided due to singularities in the amplitude-corrected waveforms at this limit.  The BBH system is parameterized as described in Sec.~\ref{sec:tidalparams}.  We use the zero detuning, high power aLIGO power spectrum \cite{PSD} for the power spectral density, and we perform inner product integrations from $f_{\rm min} = 10$ Hz to $f_{\rm max} = k F_{\rm LSO}$, where $F_{\rm LSO}$ is defined in Eq.~\eqref{eq:flso}. 

We investigate how both excluding the unphysical values of the symmetric mass ratio and including different order amplitude corrections to the waveform affect aLIGO's ability to detect deviations from the no-hair theorem expectations, as described in Sec.~\ref{sec:tidalparams}.  The bounds on the symmetric mass ratio parameter space do not decrease the minimum detectable deviation from the no-hair theorem due to the orientation of the $1\sigma$ error ellipses in the $\eta$ -- $\tilde\Lambda$ plane, as illustrated in Fig.~\ref{fig:bbhtidalplots}.   The amplitude corrections do have a noticeable affect on the measurability of tidal deformability.  While the lowest-order amplitude correction (0.5 pN) does not lead to a dramatic improvement in the measurability of the tidal parameter $\tilde \Lambda$, the 1.0 pN order amplitude correction does give about a factor of two improvement in the measurement error on $\tilde \Lambda$.  The tidal parameter is strongly correlated to both the symmetric mass ratio and the chirp mass, but these correlations are decreased, especially between chirp mass and $\tilde \Lambda$, when using the 1.0 pN order amplitude-corrected waveform.  The results are summarized in Fig.~\ref{fig:bbhtidalplots} and Table~\ref{tab:bbhtidaltable} for both the fiducial tidal parameter value of $\tilde \Lambda = 0$ and for the minimum detectable violating $\tilde \Lambda$. 

\section{Discussion}
\label{sec:discussion}

Applying physical limits on the symmetric mass ratio can have a noticeable impact on aLIGO's ability to measure spin.  When considering a near equal mass, spinning BBH system, aLIGO's ability to test the cosmic censorship conjecture is improved by about a factor of three by excluding unphysical values of the symmetric mass ratio for a Newtonian-amplitude waveform, as can be seen in Fig.~\ref{fig:spinetabound} and Table~\ref{tab:bbhspintable}.  The frequency domain waveform given in Eq.~\eqref{eq:waveform} with a Newtonian-amplitude is commonly used for detection and parameter estimation efforts.  The strong correlations between the symmetric mass ratio and spins result in poor measurability of the spin parameters when using this waveform.  However, our results imply that including a prior on the symmetric mass ratio can lead to a significant improvement of spin measurability for near equal mass, spinning BBH systems.  However, we find that a prior on the symmetric mass ratio will not affect unequal mass systems, as can be seen in Fig.~\ref{fig:nsbhspinplots}, nor will it affect near equal mass systems when amplitude-corrected waveforms are employed, as can be seen in Fig.~\ref{fig:bbhspinplots}.  

We find that switching from the Newtonian-amplitude waveform to the amplitude-corrected waveform significantly affects parameter measurability for the near equal mass, spinning BBH system, but not for the unequal mass, spinning NS-BH system.  Amplitude corrections add multiple harmonics to the gravitational waveform.  The Newtonian-amplitude waveform only includes the second harmonic.  However, the 0.5 pN order amplitude-corrected waveform adds the lowest-order point particle correction to the first and third harmonics.  The 1.0 pN order amplitude-corrected waveform adds a spin correction to the first harmonic, a point particle correction to the second harmonic and the lowest-order point particle correction to the fourth harmonic.  Parameter measurability for the spinning BBH system is most significantly affected by the 0.5 pN order point particle correction terms in the first and third harmonics and the 1.0 pN order spin correction terms in the first harmonic.  The higher order amplitude correction terms above 1.0 pN order minimally affect parameter measurability.

For the spinning BBH system, the lowest-order amplitude correction improves the measurement error for chirp mass and spin parameters by about a factor-of-ten and for the symmetric mass ratio by about a factor of fifty when compared to the Newtonian-amplitude waveform.  This translates to about a factor of two improvement on the minimum detectable spins that violate the cosmic censorship conjecture when compared to the Newtonian-amplitude waveform using the full symmetric mass ratio parameter space.  The improved measurability may be due to the breaking of the degeneracy between the symmetric mass ratio and the spin parameters.

The spinning BBH system shows significant improvement in parameter measurability again when the lowest-order spin corrections are added to the amplitude, but the spinning NS-BH system shows no significant change by including spin terms to the amplitude, as is seen in Tables~\ref{tab:bbhspintable} and~\ref{tab:nsbhspintable} and Fig.s~\ref{fig:bbhspinplots} and~\ref{fig:nsbhspinplots}.  For the spinning BBH system, the lowest-order spin corrections to the amplitude result in more than a factor-of-ten improvement in the measurability of both spin parameters when compared to the 1.0 pN order amplitude-corrected waveform without spin corrections in the amplitude.  The improved measurability may be a result of decoupling chirp mass from spin.  There is also about a factor of two improvement in the measurability of chirp mass when the lowest-order spin corrections are included in the amplitude.  Including spin corrections in the amplitude leads to about a factor of two improvement in the ability of aLIGO to detect violations of cosmic censorship for a near equal mass BBH system.

A brief study of how the 3.0 pN and 3.5 pN order spin-orbit phase corrections affect parameter measurability, summarized in Tables~\ref{tab:newspintermstableBBH} and~\ref{tab:newspintermstableNSBH}, indicates that these corrections can have a noticeable impact on the spinning BBH system and the spinning NS-BH system.  For the spinning BBH system, including the newer spin-orbit phase corrections leads to significant improvement in mass and spin measurability and a decrease in the degeneracy between spin and chirp mass for the Newtonian-amplitude waveform.  There is also some improvement in the measurability of the spin parameters for the amplitude-corrected waveforms without spin terms in the amplitude.  However, the amplitude-corrected waveforms with spin terms in the amplitude are mostly unaffected by the 3.0 pN and 3.5 pN order spin-orbit phase terms.  

The spinning NS-BH system demonstrates improved measurability for all different orders of amplitude corrections in the mass and spin parameters when the 3.0 pN and 3.5 pN order spin-orbit phase terms are included in the waveform.  More follow-up studies should be done to see how the 3.0 pN and 3.5 pN order spin-orbit phase corrections affect aLIGO's ability to detect apparent violations of the cosmic censorship conjecture.

In summary, aLIGO can theoretically detect spin violations of the cosmic censorship conjecture at $1\sigma$ for an SNR of 10 (or $3\sigma$ for an SNR of 30) for a near equal mass BBH system with component spins as small as $\chi_1 = \chi_2 = 1.16$ when using 1.0 pN order amplitude-corrected waveforms with spin corrections in the amplitude.  In addition, aLIGO can theoretically detect a spin violation at $1\sigma$ for an SNR of 10 (or $3\sigma$ for an SNR of 30) for a spinning NS-BH system with $m_1 = 1.4$ ${\rm M}_\odot$,  $m_2 = 10$  ${\rm M}_\odot$, $\chi_1 = 0$ and $\chi_2 = 4.50$ when using the 2.0 pN or 2.5 pN order amplitude-corrected waveform with spin corrections in the amplitude.

As discussed in Sec.~\ref{sec:tidalresults}, excluding unphysical values of the symmetric mass ratio does not affect aLIGO's ability to test whether the requirements of the no-hair theorem are fulfilled.  However, including amplitude corrections in the waveform does noticeably affect the measurability of the tidal deformability parameter $\tilde \Lambda$, as shown in Table~\ref{tab:bbhtidaltable} and Fig.~\ref{fig:bbhtidalplots}, which improves aLIGO's ability to detect deviations from the no-hair theorem.  There is some small improvement in parameter measurability when including the 0.5 pN order amplitude correction.  However, there is about a factor of two improvement in measurement error for both mass parameters and the tidal parameter when moving to the 1.0 pN order amplitude-corrected waveform. Note that for the nonspinning BBH system examined in Sec.~\ref{sec:tidalresults}, no spin corrections were included in the amplitude.  Therefore, the 1.0 pN order amplitude correction only adds a point particle correction to the second and fourth harmonic.  The tidal parameter $\tilde \Lambda$ is coupled to both the symmetric mass ratio and the chirp mass for waveforms including up to the 0.5 pN order amplitude correction.  The 1.0 pN order amplitude correction decouples $\tilde \Lambda$ from chirp mass and decreases the strength of the coupling between $\tilde \Lambda$ and the symmetric mass ratio.  

The minimum detectable deviation from the no-hair theorem for a near equal mass BBH system with $m_1 = 10$ ${\rm M}_\odot$ and $m_2 = 11$ ${\rm M}_\odot$ is $\tilde \Lambda = 7.61$ at $1\sigma$ for an SNR of 10 (or $3\sigma$ for an SNR of 30).  For comparison, a typical value for $\tilde \Lambda$ for a binary neutron star system is about 40, but the value of $\tilde \Lambda$ is strongly dependent on the Eq. of state \cite{BenTable}.  For an incompressible star at maximum compactness, the tidal parameter would be $\tilde \Lambda \approx 0.002$ \cite{Nagar}.

\begin{table*}[t]
	\caption{{\bf Advanced LIGO's ability to measure spin and tidal deformability, and therefore test cosmic censorship and the no-hair theorem, improves by...}}
	\bigskip
	\begin{minipage}[b]{1.0\textwidth}
	\centering
		\begin{tabular}{l l l l l}
		\hline
		& restricting $\eta$
		& including higher \hspace{0.3pt}
		&  including spin
		&  including the 3.0 pN, 3.5 pN \\
		& parameter space \hspace{0.3pt}
		& harmonics
		&  in the amplitude \hspace{0.3 pt}
		&  spin-orbit phase terms \\
		\hline
		\hline
	
		spinning BBH \hspace{0.3 pt}& yes, only & yes, starting & yes, starting & yes, mostly\\
		 & at 0.0 pN & at 0.5 pN & at 1.0 pN & at 0.0 pN \\
		 & & & & \\
		 spinning NS-BH \hspace{0.3 pt} & no & no & no & yes, for all \\
		 &  &  & & pN orders \\
		 & & & & \\
		 tidal BBH \hspace{0.3 pt} & no & yes, starting & N/A & N/A \\
		& & at 1.0 pN & & \\
		 \hline
		\end{tabular}
	\end{minipage}
	\label{tab:summary}
\end{table*}

It is worth making a brief mention of what could be causing an apparent violation of cosmic censorship or the no-hair theorem.  There could be exotic objects, such as boson stars, that do violate cosmic censorship or the no-hair theorem and therefore lead to an apparent violation through their gravitational waveform.  However, observing an apparent violation of cosmic censorship or the no-hair theorem does not necessarily mean these conjectures are false.  Rather, it could be the theory of gravity, general relativity, that is wrong, or it could be post-Newtonian theory that is wrong.  The post-Newtonian waveforms employed in this paper are based on assumptions in standard general relativity, which could be violated for systems such as a naked singularity.  However, in the case of a naked singularity, the quantum gravity effects that are fixing the singularity should only minimally affect the surrounding spacetime on which post-Newtonian waveforms are based.  In addition, the assumptions of the Kerr solution, such as axial-symmetry and asymptotic flatness, could not be satisfied.  However, detecting a nominal black hole that violates the Kerr bound or detecting internal structure in a nominal black hole would be inconsistent with the current post-Newtonian framework of general relativity and cosmological conjectures in the Kerr geometry.
 
\section{Conclusions}
\label{sec:conclusions}

We explore ways to improve aLIGO's ability to test cosmic censorship and the no-hair theorem by improving the measurability of spin and tidal deformability.  We find several methods for improving parameter measurability that affect different systems and different amplitude-corrected waveforms in different ways.  Table~\ref{tab:summary} summarizes our findings for how to improve parameter measurability for each astrophysical system that we examine.  The pN orders in the table all refer to pN order in the amplitude of the waveform, except when indicated directly.
 
Our studies indicate that both a prior on the symmetric mass ratio and including higher harmonics in the waveform can have a significant effect on aLIGO's ability to test whether expectations from the cosmic censorship conjecture and the no-hair theorem are satisfied for some, but not all, systems.  

For near equal mass spinning BBH systems, both a prior on the symmetric mass ratio and including higher harmonics could lead to significant improvement in spin and mass parameter measurability, and therefore significant improvement in aLIGO's ability to test cosmic censorship.  In addition, including spin corrections in the amplitude, specifically the lowest-order spin correction to the first harmonic, could lead to even further improved measurability of spin and mass parameters.  For the Newtonian-amplitude waveform or the waveforms with nonspinning amplitude corrections, the 3.0 pN and 3.5 pN order spin-orbit phase terms should lead to improved mass and spin measurability as well.

For the spinning NS-BH system, a prior on the symmetric mass ratio should not lead to much improvement in aLIGO's ability to test cosmic censorship.  Higher harmonics should also not improve spin or mass parameter measurability for this system.  However, the 3.0 pN and 3.5 pN order spin-orbit phase corrections should lead to improved mass and spin measurability for both Newtonian and amplitude-corrected waveforms.

For near equal mass nonspinning BBH systems with tidal corrections, a prior on the symmetric mass ratio will not improve aLIGO's ability to investigate the no-hair theorem, but including higher harmonics in the waveform will improve mass and tidal measurability.  

A final benefit of using amplitude-corrected waveforms, which include higher harmonic effects, is discussed briefly in Sec.~\ref{sec:params}.  Certain angle parameters, $\phi_c$, $\iota$, and $\phi$, are unmeasurable for a single detector with the Newtonian-amplitude waveform.  However, including the lowest-order amplitude correction in the waveform allows both $\phi_c$ and $\iota$ to become measurable, even for a single detector.  Including spin corrections in the amplitude further allows the azimuthal angle $\phi$ to become measurable for a single detector.  Therefore, higher harmonics can play a significant role in the measurability of some of the system's angle parameters, on top of the benefits to mass and spin measurability discussed above.

Overall, using a flat prior on the symmetric mass ratio and including higher harmonics in the waveform could provide aLIGO with a keen ability to test the theory of general relativity with gravitational-wave detections from black hole compact binary coalescence events in the near equal mass limit.

\section{Acknowledgments}

The authors would like to thank Leslie Wade, Richard O'Shaugnessy, Benjamin Lackey, Ian Vega, and  Badri Krishnan for helpful discussions.  We would like to thank Badri Krishnan for all of his helpful comments and time in reviewing this work.  We would also like to thank the anonymous referee for helpful comments and suggestions.  This work was supported by the NSF Grants No. PHY-0970074, No. PHY-1307429 and the Wisconsin Space Grant Consortium.

\bibliographystyle{apsrev} 
\bibliography{researchbiblio}
	
\end{document}